\documentclass{article}
\usepackage{graphicx}
\usepackage{amssymb}
\usepackage{amscd}
\usepackage{amsmath}
\usepackage[english]{babel}
\usepackage{axodraw}
\usepackage[T2A]{fontenc}
\usepackage{EuScript}
\usepackage{eufrak}
\usepackage{wrapfig}
\usepackage{float}

\newcommand{\bee}{\begin{eqnarray}}
\newcommand{\eee}{\end{eqnarray}}

\newcommand{\br}{\!\!\gg}
\newcommand{\bl}{\ll\!\!}

\def\be{\begin{eqnarray}}
\def\ee{\end{eqnarray}}
\def\nn{\nonumber}
\def\p{\partial}
\def\l{{\lambda}}
\def\b{{\beta}}
\def\a{{\alpha}}
\def\g{{\gamma}}
\def\d{{\delta}}
\def\t{{\theta}}
\def\e{{\varepsilon}}

\newcommand{\labcd}[2]{\hbox to\textwidth{#1\dotfill #2}}

\hoffset= - 1.10in         

\begin{document}

\hfill INR/TH - 24-2006

\hfill ITEP/TH - 37/06

\centerline{\Large{{\bf On Pure Spinor Superfield Formalism }}}

\bigskip
\centerline{Victor Alexandrov$^{a}$,\ \  Dmitry Krotov$^{b}$,\ \
Andrei Losev$^{c}$,\ \  Vyacheslav Lysov$^{d}$}

\begin{center}
$^a${\small{\em P.N. Lebedev Physical Institute Theoretical Physics
Division Russian Academy of Sciences, }}\\
$^b${\small{\em
Institute for Nuclear Research of the Russian Academy of Sciences, }}\\
$^{b,c,d}${\small{\em
Institute of Theoretical and Experimental Physics, }}\\
$^d${\small{\em  L.D. Landau Inst. for Theor. Phys. Russian Academy
of Sciences
,}}\\

$^{a,b}${\small{\em
Moscow State University, Department of Physics,}}\\

$^d${\small{\em Moscow Institute of Physics and Technology State
University ,}}

\end{center}
\bigskip
\centerline{ABSTRACT}
\bigskip

  We show that a certain superfield formalism can be used to find an
off-shell supersymmetric description for some supersymmetric field
theories where conventional superfield formalism does not work. This
"new" formalism contains auxiliary variables $\lambda^\alpha$ in
addition to conventional super-coordinates $\theta^\alpha$. The idea
of this construction is similar to the pure spinor formalism
developed by N.Berkovits. It is demonstrated that using this
formalism it is possible to prove that the certain Chern-Simons-like
(Witten's OSFT-like) theory can be considered as an off-shell
version for some on-shell supersymmetric field theories. We use the
simplest non-trivial model found in \cite{we} to illustrate the
power of this pure spinor superfield formalism. Then we redo all the
calculations for the case of 10-dimensional Super-Yang-Mills theory.
 The construction of off-shell description for this
theory is more subtle in comparison with the model of \cite{we} and
requires additional $Z_2$ projection. We discover {\it
experimentally} (through a direct explicit calculation) a
non-trivial $Z_2$ duality at the level of Feynman diagrams. The
nature of this duality requires a better investigation.
\bigskip

\section{Introduction}
The importance of the off-shell formulation of supersymmetric field
theories is well known. The off-shell SUSY-invariant actions can be
found only in limited number of cases for small number of
supercharges and in certain space-time dimensions. These
formulations are usually based on the superfield formalism. The main
advantage of off-shell formulation is the possibility to prove
non-renormalization theorems and derive Ward identities on
correlation functions. However, in contrast to on-shell formulation,
there are auxiliary fields in addition to physical degrees of
freedom. The number of these fields may be very large and even
infinite.

In the recent paper \cite{we} it was demonstrated that the classical
actions for different quantum field theories can be obtained as
effective actions from the single fundamental theory of Chern-Simons
(or Witten's OSFT) form
\begin{equation}\label{Fund}
S^{Fund}\ =\ \int\ Tr\Big( <\EuScript{P},\  Q_{B} \EuScript{A}>\ +\
g<\EuScript{P},\ \EuScript{A}^2>\Big)
\end{equation}
Close constructions was originally suggested in  \cite{Berkovits}
and \cite{Movshev-Schwarz}.  In the present paper we will
demonstrate that all these effective theories are in fact invariant
under the global SUSY transformation at least on-shell. The main new
results of the present consideration are that action (\ref{Fund}) is
an off-shell version of all these effective  theories and that pure
spinor formalism can be considered as a convenient superfield
formalism which allows to write the off-shell action in terms of
component fields. The definition of the fields, operator $Q_B$ and
canonical pairing $<\ ,\ >$ can be found in the section 4, see also
the introduction to \cite{we}. In this previous paper we argued that
integrating out some fields from the action (\ref{Fund}) one can
obtain physically interesting effective action. In the present paper
we show that all these fields which are integrated out are nothing
but auxiliary fields needed to restore the off-shell invariance of
the on-shell supersymmetric effective action. This view on this
procedure is very much in the spirit of \cite{Maxim}.

 The subject of the present paper is the descent of off-shell
supersymmetry of the action (\ref{Fund}) down to its effective
action. To control the SUSY properties of these actions and discuss
the descent of symmetry it is convenient to introduce an auxiliary
action $S^{SUSY}$ (interacting with superghosts) defined as
\begin{equation}\label{Fund SUSY}
S^{SUSY}\ =\ \int\ Tr\Big( <\EuScript{P},\  Q_{B} \EuScript{A}>\ +\
g<\EuScript{P},\ \EuScript{A}^2> +  <\EuScript{P},\  \e Q^s
\EuScript{A}>  +   <\EuScript{P},\ \eta^\mu \p_\mu \EuScript{A}> -
 \eta_\mu^\ast(\e\g^\mu\e)\Big)
\end{equation}
The first two terms in this action are exactly those of $S^{Fund}$.
The third and the fourth terms give the algebra of symmetry (SUSY +
translations) and the last term is determined by the structure
constants of the SUSY algebra $\{Q_\a^s, Q^s_\b\} =
2\g^\mu_{\a\b}\p_\mu$.
Here $\e^\a$ and $\eta^\mu$ are the ghosts for the global symmetry
of $S^{Fund}$ ($\varepsilon^\alpha$ - for supersymmetry, $\eta^\mu$
- for translations). Hence they do not depend on space-time
coordinates. Introduction of these auxiliary fields and addition of
the last tree terms into the action (\ref{Fund SUSY}) is needed to
guarantee that action (\ref{Fund SUSY}) satisfies classical
Batalin-Vilkovisky (BV) Master Equation over all the fields
including ghosts $\varepsilon$ and $\eta$. This fact is equivalent
to the condition that $S^{Fund}$ (the first two terms in $S^{SUSY}$)
is invariant under the algebra of symmetry generated by $Q^s$ and
$\p_\mu$ and to the condition that this algebra is closed off-shell.
As it was mentioned in the section 2 of \cite{we} (for rigorous
proof see \cite{AS}), integration of BV action over a lagrangian
submanifold preserves BV invariance of the effective action. This
invariance is what is left from the off-shell invariance of the
fundamental action. In particular it leads to the statement that the
ghost independent part of the effective action is invariant under
the on-shell  SUSY transformation.

Thus, the standard ideology of Batalin-Vilkovisky formalism allows
us to control how the off-shell symmetry of initial action is
inherited in the effective action. Usually BV formalism is used to
control gauge symmetries \cite{BV}. In this paper we apply the same
technique to study the descent of global supersymmetry.

In the section 4 we illustrate these ideas in the rather non-trivial
model with 5 quadrics found in \cite{we}. Then in section 5 we apply
the same technique  to the more interesting model  ---
10-dimensional Super Yang-Mills. Application of this procedure to
SYM is more subtle because after evaluation of effective action on
the cohomologies of $Q$-operator one should make a $Z_2$ projection
identifying the fields $\mathsf{A}$ and $\mathsf{P}$
(representatives of cohomologies in $\EuScript{A}$ and
$\EuScript{P}$). Remarkably, this projection also preserves BV
invariance  of the action. At the level of our present understanding
this fact  seems to be accidental. This observation allows us to
find  a non-trivial $Z_2$ duality  at the level of explicit
calculation of Feynman diagrams. This duality states that there are
certain identities between different Feynman diagrams like

\begin{center}
\begin{tabular}{ccl}
{\includegraphics[width=50mm]{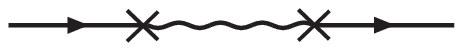}}
\put(-130,-5){$A_\mu$}\put(-20,-5){$\widetilde{c}$}\put(-105,
14){$\e Q_s$}\put(-50, 14){$\e Q_s$} &= &
{\includegraphics[width=50mm]{figure_app2.eps}}
\put(-130,-5){$c^*$}\put(-20,-6){$\widetilde{A}^*_\mu$}\put(-105,
14){$\e Q_s$}\put(-50, 14){$\e Q_s$}\\ \\$(\e\g^\mu\e)A_\mu
\widetilde{c}$ &= & $\Big(\frac3{10}+\frac9{40}+\frac14
+\frac{1}{10} + \frac3{32} - \frac3{160} +\frac{13}{320} -
\frac1{64} + \frac1{20}-\frac1{40}
\Big)(\e\g^\mu\e)\widetilde{A}^*_\mu c^*$
\end{tabular}
\end{center}
after the identification $\widetilde{c}\ =\ c^*$ and
$\widetilde{A}_\mu^*\ =\ A_\mu$. The calculation of the diagram in
the l.h.s. is almost automatic. The calculation of the diagram in
the r.h.s. is rather involved. It contains a lot of contributions.
Each contribution requires a lot of $\g$-matrix algebra including
Fiertz identities and different spinor expansions. However, all this
contributions collapse in the end  to unity which coincides with the
l.h.s. and is an example of the $Z_2$ duality. This duality has been
checked {\it experimentally} for all the diagrams arising in the
calculation of effective action. The fundamental nature of this
duality is not clear  for us. However,  what can be said is that the
action of SYM with all the SUSY structures satisfies BV equation
over all the fields as a consequence of this observed $Z_2$ duality
of Feynman diagrams.

Summarizing the introduction we would like to list our main results
obtained in the present paper.
\begin{itemize}
\item It is demonstrated that the Pure Spinor Formalism can be
considered as a Superfield Formalism for a large class of
interesting quantum field theories - pre-theories (see section 5 for definition).
\item It is shown that action (\ref{Fund}) can be considered as an
off-shell supersymmetric version of these effective pre-theories.
The action and degrees of freedom of an effective action are
dictated by the choice of the system of quadrics~$f^\mu(\l)$.
\item All effective theories  obtained after evaluation of
effective action of (\ref{Fund}) above the cohomologies of
$Q$-operator are at least on-shell supersymmetric.
\item To obtain the BV version of effective action with the SUSY
structures in case of 10-dimensional SYM one should make a $Z_2$
projection on the space of fields after the calculation of effective
action for (\ref{Fund}) (see section \ref{section SYM} for details).
For the present moment we do not know whether this $Z_2$ symmetry of
effective action can be considered  as coming from the $Z_2$
symmetry of the fundamental action. Would that be possible, we
construct the off-shell description of 10 dimensional SYM.
\item A non-trivial $Z_2$ duality  at the level of Feynman
diagrams is discovered  for the case of $SO(10)$ quadrics $f^\mu(\l)
= \l\g^\mu\l$. Evaluation of the diagrams on the one side of this
duality is almost automatic.  The corresponding calculation on the
other side is rather complicated.
\end{itemize}

\noindent Through the whole paper we use the notation
$\gamma^\mu_{\alpha \beta}$ to define the system of quadrics as
$f^\mu(\lambda)\ =\ \lambda^\alpha \gamma^\mu_{\alpha \beta}
\lambda^\beta$. We would like to stress that we do not restrict
ourselves to consider $\gamma^\mu_{\alpha \beta}$ as conventional
Dirac $\gamma$-matrices. We treat them as a set of constant
matrices, symmetric w.r.t. $\alpha$ and $\beta$. As it was shown in
\cite{we} this "extension" of the standard Berkovits' construction
allows to obtain a zoo of non-trivial effective theories for
(\ref{Fund}). Only for the case of 10-d SYM $\gamma^\mu_{\alpha
\beta}$ are conventional $SO(10)$ $\gamma$-matrices.

\subsection{From Off-shell to On-shell Theory Through BV Construction}
First of all we would like to clarify the difference between
off-shell and on-shell supersymmetric descriptions of a theory.

\noindent{\bf Off-shell description}.

\noindent Suppose there is an action $S^{cl}$ and a closed algebra
generated by $Q_\a$ and $\p_\mu$, defined by $\{Q_\a,Q_\b\} =
2\g^\mu_{\a\b}\p_\mu$. By off-shell description we mean the
following. Action $S^{cl}$ should be invariant under the
transformation
$$
\d^s_\epsilon S^{cl} = \epsilon^\a Q_\a S^{cl} = \ 0, \ \ \
\d^s_\zeta S^{cl}= \zeta^\mu \p_\mu S^{cl} = 0
$$
and that the commutator of two SUSY transformations with parameters
$\epsilon_1$ and $\epsilon_2$, being applied  to arbitrary field $A$
from the action $S^{cl}$ should satisfy
$$
[\d_{\epsilon_1},\d_{\epsilon_2}]A = 2
(\epsilon_1\g^\mu\epsilon_2)\p_\mu A
$$
In these formulas $\zeta^\mu$ is parameter for translations.

\noindent{\bf On-shell description}

\noindent By on-shell description we mean that action $S^{cl}$ is
invariant under the transformation $\d S^{cl} = 0$.  However, the
commutator of transformations, being applied to a component field
$A$ contains corrections proportional to a gauge transformation and
to the equations of motion (e.o.m.)for some fields
\be%
\label{commutator in general form}%
[\d_{\epsilon_1},\d_{\epsilon_2}]= 2
(\epsilon_1\g^\mu\epsilon_2)\p_\mu  +
\d_{gauge}+R(\epsilon_1,\epsilon_2)( \hbox{e.o.m.})
\ee%
Commutator in the l.h.s. should be applied to component fields.
 See section \ref{section Quantum Mechanics} and \ref{section
 Wess-Zumino} for details.

\noindent{\bf BV description}

A convenient tool to treat an action and symmetries on the same
footing is to use BV formalism.  The idea is to add to the classical
action $S^{cl}(\varphi)$ (here  $\varphi$ denotes all the fields)
its symmetries $V_\a (\varphi)$ with ghosts $\e^\a$ (with opposite
parity to $\epsilon^\a$) to form BV action.
\be%
 \label{BV
action Classical} S^{BV} = S^{cl} + \e^\a V_\a (\varphi) \varphi ^*
+ \eta^\mu\p_\mu\varphi\, \varphi^* + (\e\g^\mu\e)\eta^*_\mu. \ee
The fact that $S^{BV}$ satisfies BV equation is equivalent  to the
condition that $S^{cl}$ is off-shell symmetric.

The idea is to integrate the action (\ref{BV action Classical}) over
auxiliary fields. This integration preserve BV invariance. Effective
action after integration can be written as
\be%
\begin{split}
\label{effective action in general form}%
S^{eff} = S^{cl} + \e^\a V_\a (\varphi) \varphi^* +
\eta^\mu\p_\mu\varphi\, \varphi^*+ (\e\g^\mu\e)\eta^*_\mu + \\
+\text{(terms quadratic in antifields)} +\text{(terms responsible
for gauge fixing)}
\end{split}\ee%
Here the set of fields $\varphi$ in the equation (\ref{effective
action in general form}) is different from the set of fields
$\varphi$ in the equation (\ref{BV action Classical}). The same is
true for the transformations $V_\a (\varphi)$. Thus action $S^{eff}$
also satisfies BV equation. This BV action provides on-shell
description of initial theory. From the terms written in the second
line of (\ref{effective action in general form}) one can
straightforwardly extract corrections arising in the commutator
(\ref{commutator in general form}).

Thus our general philosophy can be summarized in the form of the
fig. \ref{general ideology figure}.

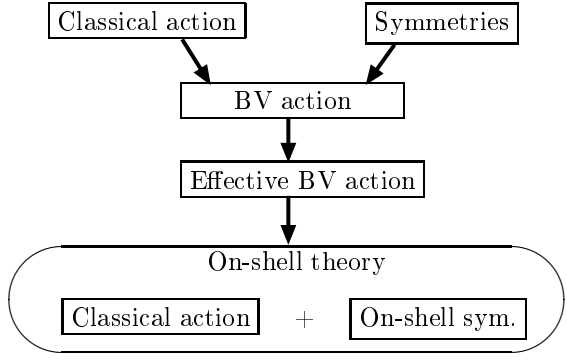
\begin{figure}[H]
 \begin{picture}(100,145)(-30,-10)

\SetScale{1}

\put(100,130){\fbox{Classical action}}
 \put(220,130){\fbox{Symmetries}}
\LongArrow(150,126)(160,110)\LongArrow(230,124)(220,110)

\put(150,100){\fbox{\ \ \ \ \ BV action \ \ \  \ }}
 \LongArrow(190,96)(190,81)

 \put(150,70){\fbox{Effective BV
  action}}
  \LongArrow(190,67)(190,50)


\put(190,28){%
 \oval(210,40) \put(-30,11){On-shell theory} \put(-85,-10){\fbox{Classical action}  \ \ \ + \ \ \ \fbox{On-shell sym.}}
}
\end{picture}%
 \caption{General philosophy}\label{general ideology
figure}
\end{figure}
First of all we unite an action and its symmetries in the  form of
BV action ($S^{SUSY}$), i.e. coupled to superghosts. Then we
integrate over lagrangian submanifold  to find effective action,
having non-standard coupling to ghosts. Finally we extract
information about on-shell theory from this effective BV action.

\section{Quantum Mechanics}
\label{section Quantum Mechanics} We start from the simplest example
- supersymmetric quantum mechanics. One can write the off-shell
SUSY-invariant action (\ref{classical action for QM}) by introducing
auxiliary field $D$. Integrating out this field one can obtain the
action which contains only physical degrees of freedom, but  is no
longer off-shell invariant. This means that the algebra of
SUSY-transformation can be closed only on-shell.

The algebra of supersymmetry is given by
\begin{eqnarray}\label{SUSY algebra}
\{Q,\bar{Q}\}=2i\partial_t \\
Q^2=\bar{Q}^2=0\nonumber
\end{eqnarray}
its representation in superspace is
\begin{eqnarray}
Q=\partial_\theta+i\bar{\theta}\partial_t\nonumber\\
\bar{Q}=\partial_{\bar{\theta}}+i\theta\partial_t\nonumber
\end{eqnarray}
and the general superfield is given by
$$
\Phi=x+\theta\bar{\psi}+\bar{\theta}\psi+\theta\bar{\theta} D
$$
The question which we discuss in the present section is: How can the
on-shell SUSY invariance  be described using BV language?

The off-shell SUSY invariant action for quantum mechanics can be
written as:
\begin{equation}\label{classical action for QM}
\begin{split}
S^{QM}\ =\
\int dt \bigg( \frac{1}{2}(\partial_tx)^2\ -\ i\bar{\psi}\partial_t
\psi\ + \ \frac{1}{2}D^2\ -\ W(x)^{\prime}D\ -\
W(x)^{\prime\prime}\psi\overline{\psi}\ \bigg)
\end{split}
\end{equation}
Here $W(\Phi)$ stands for the superpotential. The transformations of
the component fields $\d\Phi = (\e Q + \overline{\e Q})\Phi$ are
\begin{eqnarray}\label{SUSY transformation}
\delta x = \epsilon \bar{\psi}+\bar{\epsilon}\psi\nonumber \\
\delta \psi =-\epsilon (i\partial_t x+D)\\
\delta \bar{\psi} =\bar{\epsilon} (-i\partial_t x+D)\nonumber\\
\delta D=i\bar{\epsilon}\partial_t\psi-i\epsilon\nonumber
\partial_t\bar{\psi}
\end{eqnarray}
Using these expressions it is straightforward to calculate the
commutator of two supersymmetry transformations
\begin{equation}
[\delta_1,\delta_2]\psi=2i(\bar{\epsilon}_1\epsilon_2-\bar{\epsilon}_2\epsilon_1)
\partial_t\psi
\end{equation}
This result is consistent with the algebra (\ref{SUSY algebra}),
which states that anticommutator of SUSY-charges is proportional to
the shift transformation. Now we integrate upon the auxiliary field
$D$, substituting $D\ =\ W^{'}(x)$. Conducting similar computations
for the commutator one can find
\begin{equation}\label{commutator}
\begin{split}
{[\delta_1,\delta_2]}\psi\ =\
2i(\bar{\epsilon}_1\epsilon_2-\bar{\epsilon}_2\epsilon_1)
\partial_t\psi\ -\\
-(\bar{\epsilon}_1\epsilon_2-\bar{\epsilon}_2\epsilon_1)
(i\partial_t\psi-W''\psi)\ +\  2\epsilon_1\epsilon_2
(i\partial_t\bar{\psi}+W''\bar{\psi})
\end{split}
\end{equation}
The terms in the second line are proportional to the equations of
motion for the fermions. From this result it is clear that the SUSY
algebra (\ref{SUSY algebra}) is satisfied only on-shell. In a moment
we will explain how it is possible to derive these additional terms
using BV language. In the section \ref{2d gauge model} we will show
that similar terms arise after integrating out auxiliary fields in
the action (\ref{Fund}).

Let us add BV antifields and ghosts to the classical action for SUSY
QM
\begin{equation}\label{initial BV action for SUSY-QM}
\begin{split}
S^{\scriptscriptstyle BV}\ =\ \int\ \frac{1}{2}(\partial_t x)^2\ +\
\frac{1}{2}D^2\ -\ i\bar{\psi}\partial_t \psi\ -\ (W^{\prime}D\  +\
W^{\prime\prime}\psi\overline{\psi})\ +\\ +\ (\varepsilon
\bar{\psi}+\bar{\varepsilon}\psi)x^*\  -\ \varepsilon (i\partial_t
x+D)\psi^*\ +\ \bar{\varepsilon} (-i\partial_t x+D)\bar{\psi}^*\ +\
(i\bar{\varepsilon}\partial_t\psi-i\varepsilon
\partial_t\bar{\psi})D^*\ +\\+\ \eta\partial_t x x^*\  +\ \eta\partial_t \psi \psi^*\
+\ \eta\partial_t\bar{\psi}\bar{\psi}^*\ +\ \eta\partial_t D D^*\ +\
2i \varepsilon \bar{\varepsilon}\eta^*
\end{split}
\end{equation}
The first line of this expression is the classical action for SUSY
QM, the second one is BV structure of SUSY transformation (see
(\ref{SUSY transformation}) for the transformations of the component
fields), the last line contains BV structure for translations in
time (this is necessary to close the algebra of symmetry) and the
term with the structure constants for the symmetry algebra (the last
term). We would like to emphasize the difference between parameter
$\epsilon^\alpha$ of SUSY transformation (see for example
(\ref{commutator}) ) and the ghost $\varepsilon_\alpha$ for SUSY
transformation used in (\ref{initial BV action for SUSY-QM}). The
first one is odd variable, the second one is even. There is complete
analogy with gauge theories here: parameter of gauge transformation
is even, while parameter of BRST transformation (Faddeev-Popov
ghost) is odd. The parity of the ghost for the transformation is
always opposite to the parity of the parameter. The ghost field for
translations is denoted by $\eta$ (odd variable), the BV anti-ghost
for translations is $\eta^\ast$ (even variable).

  Action (\ref{initial BV action for SUSY-QM}) satisfies classical BV equation:
$$
\int\ \frac{\delta_{\scriptscriptstyle L} S}{\delta
\chi}^{\!\scriptscriptstyle BV}\!\frac{\delta_{\scriptscriptstyle R}
S}{\delta \chi^\ast}^{\!\scriptscriptstyle BV}\  = \ \int\
\frac{\delta_{\scriptscriptstyle L} S}{\delta
x}^{\!\scriptscriptstyle BV}\!\frac{\delta_{\scriptscriptstyle R}
S}{\delta x^\ast}^{\!\scriptscriptstyle BV}\ +\ \
\frac{\delta_{\scriptscriptstyle L} S}{\delta
\psi}^{\!\scriptscriptstyle BV}\!\frac{\delta_{\scriptscriptstyle R}
S}{\delta \psi^\ast}^{\!\scriptscriptstyle BV}\ +\ \
\frac{\delta_{\scriptscriptstyle L} S}{\delta
\bar{\psi}}^{\!\scriptscriptstyle
BV}\!\frac{\delta_{\scriptscriptstyle R} S}{\delta
\bar{\psi}^\ast}^{\!\scriptscriptstyle BV}\ +\ \
\frac{\delta_{\scriptscriptstyle L} S}{\delta
D}^{\!\scriptscriptstyle BV}\!\frac{\delta_{\scriptscriptstyle R}
S}{\delta D^\ast}^{\!\scriptscriptstyle BV}\ +\ \
\frac{\delta_{\scriptscriptstyle L} S}{\delta
\eta}^{\!\scriptscriptstyle BV}\!\frac{\delta_{\scriptscriptstyle R}
S}{\delta \eta^\ast}^{\!\scriptscriptstyle BV}\ =\ 0
$$
Here $\chi$ stands for all the fields. There are no terms arising
from the variation over $\varepsilon$ and $\overline{\varepsilon}$
because the action $S^{\scriptscriptstyle BV}$ is independent of the
antifields $\varepsilon^\ast$ and $\overline{\varepsilon}^\ast$.
Integrating BV action over a lagrangian submanifold results into
effective action which again satisfies BV equation (the simplistic
explanation of this fact is given in the section 2 of \cite{we}, for
the rigorous proof see \cite{AS}). Let us integrate over the
auxiliary field $D$ on the lagrangian submanifold $D^\ast\ =\ 0$.
The result for the effective action is:
\begin{equation}\label{QM effective action}
\begin{split}
S^{eff}\ =\ \int\ \frac{1}{2}(\partial_t x)^2\ -\
i\bar{\psi}\partial_t \psi\ -\ \frac{1}{2}W^{\prime 2}\ -\
W^{\prime\prime}\psi \bar{\psi}\ +\\ +\ (\varepsilon
\bar{\psi}+\bar{\varepsilon}\psi)x^*\  -\ i\varepsilon \partial_t
x\psi^*\ -\ i\bar{\varepsilon} \partial_t x\bar{\psi}^*\  -\
W^{\prime}(\varepsilon \psi^\ast\ -\
\bar{\varepsilon}\bar{\psi}^\ast)\ +\\ +\ \eta\partial_t x x^*\ +\
\eta\partial_t \psi \psi^*\ +\ \eta\partial_t\bar{\psi}\bar{\psi}^*\
+\ 2i \varepsilon \bar{\varepsilon}\eta^*\ -\\ -\
\frac{1}{2}(\varepsilon\psi^\ast\ -\
\bar{\varepsilon}\bar{\psi}^\ast)^2
\end{split}
\end{equation}
Similarly to (\ref{initial BV action for SUSY-QM}) in the first line
we have classical action, in the second one - SUSY transformations
of the remaining fields, the third line contains BV structure for
translations and the structure constants term.

The most interesting term appears in the last line. It is quadratic
in the antifields and quadratic in the ghosts $\varepsilon$ and
$\bar{\varepsilon}$. Let us decompose the effective action $S^{eff}$
into two parts $S^{eff}\ =\ S^{}_{_S}\ +\ S^{add}$, where $S^{add}\
=\ -\frac{1}{2}(\varepsilon\psi^\ast\ -\
\bar{\varepsilon}\bar{\psi}^\ast)^2$. Subscript "$S$" in $S^{}_{_S}$
denotes the action with standard (linear in antifields) coupling to
ghosts. Since (\ref{QM effective action}) is obtained from the
integration of BV action over the lagrangian submanifold, $S^{eff}$
satisfies classical BV equation, which can be written as
\begin{equation}
0\ =\ \int\ \frac{\delta_{\scriptscriptstyle L} S}{\delta
\chi^n}^{\!eff}\!\!\frac{\delta_{\scriptscriptstyle R}
S}{\delta\chi_n^\ast}^{\!eff}\ =\ \ \int\
\frac{\delta_{\scriptscriptstyle L} S_{_S}}{\delta \chi^n}^{\ \
}\!\!\frac{\delta_{\scriptscriptstyle R}
S_{_S}}{\delta\chi_n^\ast}^{\ \ }\ +\
\frac{\delta_{\scriptscriptstyle L} S_{_S}}{\delta \psi}^{\ \
}\!\!\frac{\delta_{\scriptscriptstyle R}
S}{\delta\psi^\ast}^{\!\!add}\ +\ \frac{\delta_{\scriptscriptstyle
L} S_{_S}}{\delta \bar{\psi}}^{\ \
}\!\!\frac{\delta_{\scriptscriptstyle R}
S}{\delta\bar{\psi}^\ast}^{\!\!add}
\end{equation}
Taking into account explicit expressions for $S^{add}$ and
$S^{}_{_S}$ one can rewrite this result as
\begin{equation}\label{BV invariance of the total action}
0\ =\ \int\ \frac{\delta_{\scriptscriptstyle L} S}{\delta
\chi^n}^{\!eff}\!\!\frac{\delta_{\scriptscriptstyle R}
S}{\delta\chi_n^\ast}^{\!eff}\ =\ \ \int\
\frac{\delta_{\scriptscriptstyle L} S_{_S}}{\delta \chi^n}^{\ \
}\!\!\frac{\delta_{\scriptscriptstyle R}
S_{_S}}{\delta\chi_n^\ast}^{\ \ }\ +\
\frac{\delta_{\scriptscriptstyle L} S}{\delta
\psi}^{\!QM}\!\!\Big(\varepsilon\bar{\varepsilon}\bar{\psi}^\ast\ -\
\varepsilon^2\psi^\ast\Big)\ +\ \frac{\delta_{\scriptscriptstyle L}
S}{\delta
\bar{\psi}}^{\!QM}\!\!\Big(\varepsilon\bar{\varepsilon}\psi^\ast\ -\
\bar{\varepsilon}^2\bar{\psi}^\ast\Big)
\end{equation}
where $S^{QM}$ is the classical action for quantum mechanics
(\ref{classical action for QM}). One can see that the last two terms
vanish on the equations of motion for the fields $\psi$ and
$\bar{\psi}$. Reducing equation (\ref{BV invariance of the total
action}) to the solutions of the classical equations of motion one
can obtain
$$
\int\ \frac{\delta_{\scriptscriptstyle L} S_{_S}}{\delta \chi^n}^{\
\ }\!\!\frac{\delta_{\scriptscriptstyle R}
S_{_S}}{\delta\chi_n^\ast}^{}\Bigg|_{\hbox{on the e.o.m.}} = \ \ \ 0
$$
which is the condition of SUSY invariance. This condition however is
valid only on-shell.

 What is important for us from this calculation is that appearance
of $\varepsilon^2(\chi^{*})^2$ terms in the effective action signals
the descent of off-shell invariance of the fundamental action down
to on-shell invariance of the effective action.

\subsection{General case}
Though this effect was illustrated using the simplest possible
example --- supersymmetric QM interpretation of these terms
(quadratic in antifields and in the ghosts for SUSY) is universal
and does not depend on the particular theory. To demonstrate this
one can write the general structure of effective action as
\begin{equation}%
\label{general structure for eff theory}%
S^{eff}\ =\ S^{cl}\ +  c ^\alpha V^n_\alpha \chi^*_n -\
\frac{1}{2}f^\gamma_{\alpha \beta }  c  ^\alpha  c  ^\beta
 c _\gamma^* + S^{add},
\end{equation}%
where we used the notation $V_\alpha^n = Q_\alpha \chi^n$  in the
term describing the transformation of the component fields. The
algebra of symmetry  is given by $[ Q_\alpha, Q_\beta  ] =
f^\gamma_{\alpha \beta } Q_\gamma$ and $c^\alpha$ are ghosts for
this algebra. One can straightforwardly plug this action into the
classical BV equation and collect the terms linear  in antifields.
The result is given by
\be%
\label{formula 1}%
 c ^\alpha V_\alpha^k  c ^\beta \frac{\delta
V^n_\beta}{\delta \chi^k}\chi^*_n -  \frac12 V^n_\alpha \chi^*_n
f^\alpha_{\beta\gamma} c ^\beta  c ^\gamma + \frac{\delta
S^{cl}}{\delta \chi^k} \frac{\delta S^{add}}{\delta \chi_k^*}= 0.
\ee%
Varying this expression  w.r.t. $ c ^\alpha,  c ^\beta$ and
$\chi^*_n$ one can find

\be%
\label{formula 2}%
 V_\alpha^k
\frac{\delta V_\beta^n}{\delta \chi^k} + V^k_\beta \frac{\delta
V_\alpha^n}{\delta \chi^k} -f^\gamma_{\alpha\beta} V^n_\gamma +
\frac{\delta S^{cl} }{\delta \chi^k}
\frac{\delta^{(4)}S^{add}}{\delta  c ^\alpha \  \delta
 c ^\beta \ \delta\chi^*_k\ \delta\chi^*_n} = 0,
\ee%
which is equivalent to
\be%
\label{formula 3}%
 [Q_\alpha, Q_\beta  ] \chi^n  = f^\gamma_{\alpha\beta} Q_\gamma
 \chi^n -
\frac{\delta S^{cl} }{\delta \chi^k}
\frac{\delta^{(4)}S^{add}}{\delta  c ^\alpha \  \delta
 c ^\beta \ \delta\chi^*_k\ \delta\chi^*_n} = 0,
\ee%
and coincides with (\ref{commutator}): the  commutator of $\d_1$ and
$\d_2$ is connected with the commutator of $Q$.

Thus the non-standard terms (quadratic in antifields)  in BV action
are in one-to-one correspondence with the corrections proportional
to the equations of motion (\ref{commutator}). This result is not
new. For the non-complete list of references on the subject see
\cite{Hull}.

\section{Wess-Zumino Gauge}
\label{section Wess-Zumino}In this section we realize the gauge
fixing procedure using the BV language in the simple and well known
example: Wess-Zumino gauge in $\mathcal{N}=1$ four dimensional
super-Maxwell theory. We find non-standard terms in the solution of
BV Master Equation responsible for the fact that the Wess-Zumino
gauge is not supersymmetric (the SUSY transformation of the vector
multiplet in the Wess-Zumino gauge gives the fields which are absent
in this gauge; to restore the Wess-Zumino gauge one should make an
appropriate gauge transformation). In the next section we will
demonstrate that exactly these terms appear in the effective action
for (\ref{Fund}) after evaluation on the cohomologies of $Q$
operator. This observation will lead to the conclusion that the
action (\ref{Fund}) contains the full multiplet of auxiliary fields
needed to restore the SUSY invariance, while in the effective action
these fields are integrated out.

\subsection{Gauge-fixing procedure in BV description}\label{gauge fixing in BV
description} In this subsection we will show that restriction of the
Master Action of BV formalism to the certain lagrangian submanifold
gives Faddeev-Popov  action in the fixed gauge. As an example
consider the BV action for the gauge invariant action $S^{cl}$ which
depends only on the gauge field (no matter fields):
\begin{equation}\label{BV action for the gauge theory}
S^{BV}\ =\ S^{cl}\ +\ \int D_\mu^{a c}c^c(A_\mu^a)^\ast\ -\
\frac{1}{2}f^a_{b c}c^bc^c(c_a)^\ast
\end{equation}
We are going to restrict this action to the certain lagrangian
submanifold $\EuScript{L}_f$. The definition of $\EuScript{L}_f$ is
given by:
\begin{equation}\label{lagrangian sub-manifold L^ast}
\left\{\begin{array}{l}
(c^a)^\ast\ =\ 0\\
f^a(A)\ =\ 0 \\
(A_\mu^a)^\ast\ =\ -\frac{\partial f^b}{\partial
A_\mu^a}\overline{c}^b \\
c^a\  -\  $\footnotesize{is not restricted}$
\end{array}\right.
\end{equation}
Thus coordinates on this submanifold are $c^a$, $\overline{c}^a$ and
$A_\mu^a$ restricted by the constraint $f^a(A) = 0$ . The first
coordinate $c^a$ is not restricted, while there is the constraint
$f^a(A)\ =\ 0$, imposed on the field $A_\mu^a$. The additional
degree of freedom appearing  in the field $(A_\mu^a)^\ast$ is
parameterized by $\overline{c}^b$.  As we will show in a moment this
coordinate $\overline{c}^b$ on the submanifold is nothing but
Faddeev-Popov antighost field. It is straightforward to check that
(\ref{lagrangian sub-manifold L^ast}) is indeed a lagrangian
submanifold: \vspace{-0.3cm}\begin{equation}\nonumber
\begin{split}
\delta\chi_n^\ast\wedge\delta\chi^n\ =\ \delta c^\ast\wedge\delta c\
+\ \delta(A_\mu^a)^\ast\wedge\delta A_\mu^a\ =\ \delta
A_\mu^a\wedge\delta\left(\frac{- \partial f^b}{\partial
A_\mu^a}\overline{c}^b\right)\ =\ \ \ \ \ \ \ \ \ \ \ \ \ \\ =\
-\delta A_\mu^a\wedge\frac{\partial^2f^b}{\partial A_\nu^c\partial
A_\mu^a}\overline{c}^b\delta A_\nu^c\ -\ \delta
A_\mu^a\wedge\delta\overline{c}^b\frac{\partial f^b}{\partial
A_\mu^a}\ =\ \frac{\partial^2f^b}{\partial A_\nu^c\partial
A_\mu^a}\overline{c}^b\delta A_\mu^a\wedge\delta A_\nu^c\ -\ \delta
f_b\wedge\delta\overline{c}^b\ =\ 0
\end{split}
\end{equation}
We start with BV symplectic form written for all the fields and
antifields of the theory. In the second equality we used
$(c^a)^\ast\ =\ 0$ and plugged $(A_\mu^a)^\ast$ from
(\ref{lagrangian sub-manifold L^ast}). In the next equality we apply
operator $\delta$ to $\frac{\partial f_b}{\partial A_\mu^a}$ and to
$\overline{c}^b$. The first term in the last equality vanishes
because $\frac{\partial^2 f_b}{\partial A_\nu^c\partial A_\mu^a}$ is
symmetric under interchange $A_\mu^a\ \leftrightarrow\ A_\nu^c$
while $\delta A_\mu^a\wedge\delta A_\nu^c$ is antisymmetric. The
second term is equal to zero because of the constraint $f^b\ =\ 0$
in (\ref{lagrangian sub-manifold L^ast}).

  Restricting the action (\ref{BV action for the gauge theory}) to
the lagrangian submanifold (\ref{lagrangian sub-manifold L^ast}) one
can obtain
$$
S^{BV}\bigg|_{\EuScript{L}_f}\ = \!\!\!= \ S^{cl}\ +\ \int
D_{\mu}^{a c} c^c (A_{\mu}^a)^{\ast}\ -\ \frac{1}{2}g f^{a}_{\ b c}
c^b c^c (c^{a})^\ast\ \bigg|_{\EuScript{L}_f} =\ S^{cl}\ -\ D_\mu^{a
c}c^c\frac{\partial f_b}{\partial A_\mu^a}\overline{c}^b\ \
\bigg|_{f_b(A)\ =\ 0}
$$
which is Faddeev-Popov action in the fixed gauge and the coordinate
on the lagrangian submanifold $\bar{c}^b$ is Faddeev-Popov antighost
field.

Summarizing this calculation one can see that to fix the certain
gauge one should plug the gauge restriction on the fields into the
action, integrate over the corresponding antifield, introducing the
coordinate on the lagrangian submanifold according to
(\ref{lagrangian sub-manifold L^ast}), and put the BV antifield for
the ghost equal to zero, integrating over the ghost.

\subsection{Gauge-fixing of the Wess-Zumino gauge}
We start from a set of definitions for the SUSY multiplets. We use
the standard 2-d notations for the superfields (see  for example
\cite{Bilal}). In these notations the chiral multiplet is given by
\begin{equation}
\begin{split}
\Lambda\ =\ (a+ib)(y)\ +\ \theta\psi(y)\ -\ \theta\theta F(y)\ =\ \
\
\ \ \ \ \ \ \ \ \ \ \ \ \ \ \ \ \ \ \ \ \ \ \ \ \ \ \ \ \ \\
=\ (a+ib)\ +\ \theta\psi\ +\
i\theta\sigma^\mu\bar{\theta}\partial_\mu(a+ib)\ -\ \theta\theta F\
-\ \frac{i}{2}\theta\theta\partial_\mu\psi\sigma^\mu\bar{\theta}\ -\
\frac{1}{4}\theta\theta\overline{\theta\theta}\partial^2(a+ib)
\end{split}
\end{equation}
the contraction of indices is given by $\theta\theta\ =\
\theta^\alpha\theta_\alpha$, while $\overline{\theta\theta}\ =\
\bar{\theta}_{\dot{\alpha}}\bar{\theta}^{\dot{\alpha}}$ and $y^\mu\
=\ x^\mu\ +\ i\theta\sigma^\mu\bar{\theta}$. The SUSY variation of
component fields can be found by direct application of SUSY charges:
\begin{eqnarray}
Q_\alpha\ =\ \frac{\partial}{\partial\theta^\alpha}\ -i\sigma^\mu_{\alpha\dot{\beta}}\bar{\theta}^{\dot{\beta}}\partial_\mu\\
\bar{Q}_{\dot{\alpha}}\ =\
-\frac{\partial}{\partial\bar{\theta}^{\dot{\alpha}}}\ +\
i\theta^\beta\sigma^\mu_{\beta\dot{\alpha}}\partial_\mu\nonumber
\end{eqnarray}
(here $\partial_\mu$ denotes the derivative w.r.t. $x^\mu$)
according to the rule
$$
\delta\Lambda\ =\ (\epsilon Q\ +\ \overline{\epsilon Q}) \Lambda
$$
Performing simple calculations one can find:
\begin{eqnarray}\label{component SUSY trans chiral mult}
\delta a\ =\ \frac{1}{2}(\epsilon\psi\ +\
\overline{\epsilon\psi})\nonumber\\
\delta b\ =\ \frac{i}{2}(\overline{\epsilon\psi}\ -\
\epsilon\psi)\\
\delta\psi\ =\ -2\epsilon F\ -\
2i\partial_\mu(a+ib)\ \bar{\epsilon}\bar{\sigma}^\mu\nonumber\\
\delta F\ =\
-i\bar{\epsilon}\bar{\sigma}^\mu\partial_\mu\psi\nonumber
\end{eqnarray}
Similar computation for the vector multiplet
\begin{equation}
\begin{split}
V(x,\theta,\bar{\theta})\ =\ C\ +\ i\theta\chi\ -\
i\overline{\theta\chi}\ +\ \theta\sigma^\mu\bar{\theta}A_\mu\ +\
\frac{i}{2}M\theta\theta\ -\
\frac{i}{2}\overline{M}\overline{\theta\theta}\ +\\+\
i\theta\theta\big( \overline{\theta\lambda}\ +\
\frac{i}{2}\bar{\theta}\bar{\sigma}^\mu\partial_\mu\chi\big)\ -\
i\overline{\theta\theta}\big(\theta\lambda\ +\
\frac{i}{2}\theta\sigma^\mu\partial_\mu\bar{\chi}\big)\ +\
\frac{1}{2}\theta\theta\overline{\theta\theta}\big(D\ -\
\frac{1}{2}\partial^2 C\big)
\end{split}
\end{equation}
gives the following component transformations:
\begin{eqnarray}\label{component SUSY trans vector mult}
\delta C\ =\ i\epsilon\chi\ -\ i\overline{\epsilon\chi}\nonumber\\
\delta\chi\ =\ \epsilon M\ -\ \partial_\mu C\
\bar{\epsilon}\bar{\sigma}^\mu\ +\
iA_\mu\ \bar{\epsilon}\bar{\sigma}^\mu\nonumber\\
\delta A_\mu\ =\ -(\epsilon\partial_\mu\chi)\ +\
i(\epsilon\sigma^\mu\bar{\lambda})\ -\
\overline{\epsilon\partial_\mu\chi}\ +\
i\bar{\epsilon}\bar{\sigma}^\mu\lambda\\
\delta\lambda\ =\ i\epsilon D\ -\
\frac{1}{2}\epsilon(\sigma^\nu\bar{\sigma}^\mu\ -\
\sigma^\mu\bar{\sigma}^\nu)\partial_\mu A_\nu\nonumber\\
\delta D\ =\ -\epsilon\sigma^\mu\partial_\mu\bar{\lambda}\ +\
\bar{\epsilon}\bar{\sigma}^\mu\partial_\mu\lambda\nonumber\\
\delta M\ =\ 2\overline{\epsilon\lambda}\ +\
2i(\bar{\epsilon}\bar{\sigma}^\mu\partial_\mu\chi)\nonumber
\end{eqnarray}
It is well known that the gauge transformation for the vector
multiplet is given by:
$$
V\ \longrightarrow\ V\ +\ \frac{1}{2} (\Lambda\ +\ \bar{\Lambda})
$$
which in component fields gives:
\begin{eqnarray}\label{component gauge transformation}
\delta C\ =\ a\nonumber\\
\delta\chi\ =\ -\frac{i}{2}\psi\nonumber\\
\delta\bar{\chi}\ =\ \frac{i}{2}\bar{\psi}\\
\delta A_\mu\ =\ -\partial_\mu b\nonumber\\
\delta M\ =\ -\bar{F},\ \ \ \ \ \ \ \delta\bar{M}\ =\ -F\nonumber\\
\delta\lambda\ =\ \delta\bar{\lambda}\ =\ \delta D\ =\ 0\nonumber
\end{eqnarray}
Now we are going to demonstrate how it is possible to fix the
Wess-Zumino gauge in the abelian super Maxwell theory using BV
formalism. We will show that after the gauge fixing some
non-standard terms appear in the BV action. These terms are
responsible for the fact that the Wess-Zumino gauge is not
supersymmetric - commutator of two SUSY transformations, should be
accompanied by the appropriate gauge transformation to return into
the Wess-Zumino gauge. As before we introduce the ghosts for the
SUSY transformations: $\varepsilon$ and $\bar{\varepsilon}$ and the
ghosts for translations $\eta^\mu$. The full Master Action of BV
formalism can be schematically written as:
\begin{equation}
\begin{split}
S^{BV}\ =\ \int\ -\frac{1}{4}F_{\mu\nu}^2\ -\
i\lambda\sigma^\mu\partial_\mu\bar{\lambda}\ +\ \frac{1}{2}D^2\ +
\ \  \ \ \ \ \ \  \ \ \ \ \ \ \ \ \ \ \ \ \ \ \ \ \ \ \ \ \ \ \  \ \ \ \\
+\ \frac{1}{2}(\Lambda\ +\ \bar{\Lambda})V^\ast\ +\ (\varepsilon Q\
+\ \overline{\varepsilon Q}\ +\ \eta^\mu\partial_\mu)V\ \! V^\ast\
+\ (\varepsilon Q\ +\ \overline{\varepsilon Q}\ +\
\eta^\mu\partial_\mu)\Lambda\ \! \Lambda^\ast\ -\
2i\eta^\ast_\mu(\varepsilon\sigma^\mu\bar{\varepsilon})
\end{split}
\end{equation}
In the first line of this expression the classical abelian gauge
invariant action is written. The second line contains the gauge and
SUSY transformation of all the fields as well as the structure
constant term resulting from the anticommutator of the SUSY charges
(\ $\{Q_\alpha,\ \bar{Q}_{\dot{\alpha}}\}\ =\
2i\sigma^\mu_{\alpha\dot{\alpha}}\partial_\mu$\ ). To write explicit
expressions in the component fields one should take the component
transformations from (\ref{component SUSY trans chiral mult}),
(\ref{component SUSY trans vector mult}), (\ref{component gauge
transformation}) and multiply them by the appropriate antifield. For
example one of the terms resulting from $\frac{1}{2}(\Lambda\ +\
\bar{\Lambda})V^\ast$ gives $-\partial_\mu b\ \! (A_\mu)^\ast$ (see
the fourth line of (\ref{component gauge transformation})\ ).

\noindent{\bf WARNING.} The fields of chiral multiplet are ghosts
for the gauge transformation. Hence their parities are opposite to
the standard parities of the component fields in the chiral
multiplet. The fields $a$, $b$, $F$, $\bar{F}$ are odd, while $\psi$
and $\bar{\psi}$ are even.

Now we are going to fix the Wess-Zumino gauge $C\ =\ 0$, $\chi\ =\
0$, $M\ =\ 0$ using the procedure discussed in the subsection
\ref{gauge fixing in BV description}. To do this one should put the
fields: $C$, $\chi$, $\bar{\chi}$, $M$, $\bar{M}$ equal to zero as
well as antighosts: $a^\ast$, $\psi^\ast$, $\bar{\psi}^\ast$,
$F^\ast$, $\bar{F}^\ast$ and integrate over the antifields $C^\ast$,
$\chi^\ast$, $\bar{\chi}^\ast$, $M^\ast$, $\bar{M}^\ast$ as well as
over the ghosts $a$, $\psi$, $\bar{\psi}$, $F$, $\bar{F}$. Direct
computation taking into account the parities of all the fields
gives\footnote{Here we denote the gauge ghost by $b$ to avoid
confusion with the first component $C$ of the vector multiplet.}
\begin{equation}
\begin{split}
S^{eff} =\int \Big[-\frac{1}{4}F_{\mu\nu}^2 -
i\lambda\sigma^\mu\partial_\mu\bar{\lambda}\ +\ \frac{1}{2}D^2\ -
\partial_\mu b (A_\mu)^\ast\ +\ i(\varepsilon \sigma^\mu\bar{\lambda}\ +\
\lambda\sigma^\mu\bar{\varepsilon}) A_\mu^\ast\ -\
\big(\varepsilon\sigma^\mu\partial_\mu\bar{\lambda}\ -\
\partial_\mu\lambda\sigma^\mu\bar{\varepsilon}\big) D^\ast\\ +\
\Big(i\varepsilon D\ -\
\frac{1}{2}\varepsilon(\sigma^\mu\bar{\sigma}^\nu\ -\
\sigma^\nu\bar{\sigma}^\mu) \partial_\mu A_\nu\Big)\lambda^\ast\ +\
\Big(-i\bar{\varepsilon} D\ -\
\frac{1}{2}(\bar{\sigma}^\nu\sigma^\mu\ -\
\bar{\sigma}^\mu\sigma^\nu)\bar{\varepsilon} \partial_\mu
A_\nu\Big)\bar{\lambda}^\ast\
 +\\ +\
\big(\eta^\mu\partial_\mu A^\nu A_\nu^\ast\ +\ \eta^\mu\partial_\mu
D D^\ast\ +\ \eta^\mu\partial_\mu \lambda \lambda^\ast\ +\
\eta^\mu\partial_\mu \bar{\lambda} \bar{\lambda}^\ast\ +\
\eta^\mu\partial_\mu b b^\ast\ \big)\ -\ 2i(\varepsilon\sigma^\mu\bar{\varepsilon})\eta_\mu^\ast\ -\\
-\ 2i(\varepsilon\sigma^\mu\bar{\varepsilon})A_\mu b^\ast\ \Big]
\end{split}
\end{equation}
The aim of this  calculation  was to demonstrate  the appearance of
the last term $2i\varepsilon\sigma^\mu\bar{\varepsilon}A_\mu b^*$.
Following the logic mentioned in the end of the previous section one
can check  that these terms  are responsible for the fact that the
gauge which is fixed is {\it not consistent} with the supersymmetry.
The algebra of SUSY is closed only up to the gauge transformation
with parameter $(\varepsilon\sigma^\mu\bar{\varepsilon})A_\mu$. In
the next two sections we will show, that the terms discussed  in the
sections 2 and 3 are exactly those which arise when one integrates
out auxiliary fields in the action (\ref{Fund}) to obtain effective
action. This will be shown for the model found in \cite{we} and for
Berkovits' 10-dimensional SYM theory.

\

\

\section{Superfield Formulation of Gauge Model of \cite{we}}\label{2d gauge model}
In \cite{we} we introduced  the model that is believed to be the
simplest example in the class of physically interesting models.
Despite of its own interest this model can be considered as a toy
model which inherits almost all the phenomena related to the descent
of supersymmetry in case of 10-dimensional SYM, which is the main
subject of the present paper. In \cite{we} it was demonstrated that
the classical part of effective BV action calculated for the theory
(\ref{Fund}) in case the system of quadratic constraints
$f^{\mu}(\lambda)$ is given by:
\begin{eqnarray}\label{singular quadrics}
f_1\ =\ \lambda_1\lambda_2\nonumber\\
f_2\ =\ \lambda_2\lambda_3\nonumber\\
f_3\ =\ \lambda_3\lambda_4\\
f_4\ \ \ =\ \ \lambda_1^2\nonumber\\
f_5\ \ \ =\ \ \lambda_4^2\nonumber
\end{eqnarray}
can be written as:
\begin{equation}\label{classical action for 2d gauge model}
\begin{split}
S^{cl} = \int\!\!d^2x\ Tr\ \!\bigg( \Phi F_{+ -} +
D_+\phi_1D_-\phi_1 + D_-\phi_2D_+\phi_2 -
\frac{g}{\sqrt{2}}\phi_1\{\psi_+,\psi_-\} +
i\frac{g}{\sqrt{2}}\phi_2\{\psi_+,\psi_-\} + \beta_+D_-\gamma_+ +\\
+ \beta_-D_+\gamma_- + \overline{\psi}_-D_+\psi_- +
\overline{\psi}_+D_-\psi_+ + \overline{\chi}_-D_+\chi_- +
\overline{\chi}_+D_-\chi_+ + 2g\overline{\chi}_-[\gamma_-,\psi_+] +
2g\overline{\chi}_+[\gamma_+,\psi_-] \bigg)
\end{split}
\end{equation}
The aim of this section is to explain that this action is invariant
under the global supersymmetry transformation on-shell in the same
sense that  $\mathcal{N}=1$ Yang-Mills action in the Wess-Zumino
gauge is invariant under the supersymmetry transformation (the
action is invariant and the SUSY algebra is closed up to an
appropriate gauge transformation). Another point is that action
(\ref{Fund}) is the off-shell version of the theory (\ref{classical
action for 2d gauge model}) without elimination of auxiliary fields
like $C$, $\psi$, $M$ in the Wess-Zumino gauge.

\subsection{Initial BV action}
 Firstly we notice that there is an odd supersymmetry generator,
built using the quadrics $f^\mu(\lambda)$, which anticommutes with
the Berkovits operator
\begin{equation}\label{QBerkovits}
Q_{B}\ =\ Q\ +\ \Phi\ =\
\lambda_\alpha\frac{\partial}{\partial\theta_\alpha}\ +\
\theta_{\alpha}\frac{\partial
f^{\mu}}{\partial\lambda_\alpha}\partial_{\mu}
\end{equation}
This SUSY generator is given by:
\begin{equation}
Q^{\scriptscriptstyle SUSY}_\alpha \ =\
\frac{\partial}{\partial\theta^\alpha}\ -\
\frac{1}{2}\theta_\beta\frac{\partial^2
f^\mu}{\partial\lambda_\beta\partial\lambda_\alpha}\partial_\mu
\end{equation}
By straightforward calculation one can see that
$$
\{\ Q_B\ ,\ Q^{\scriptscriptstyle SUSY}_\alpha\}\ =\ \frac{\partial
f^\mu}{\partial\lambda_\alpha}\partial_\mu\ -\
\frac{1}{2}\lambda_\beta\frac{\partial^2
f^\mu}{\partial\lambda_\beta\partial\lambda_\alpha}\partial_\mu\ =\
0
$$ This is true, because the functions $f^\mu(\lambda)$ are
quadratic in $\lambda_\alpha$. Substituting explicit expressions for
quadrics $f^\mu(\lambda)$ one can find the following expressions for
the supersymmetry generators(we omit the superscripts SUSY):
$$
{\begin{array}{l} Q_1\ =\ \frac{\partial}{\partial\theta_1}\ -\
2\theta_1\partial_+\ \ \ \ \ \ \ Q_2\ =\
\frac{\partial}{\partial\theta_2}
\\Q_4\ =\ \frac{\partial}{\partial\theta_4}\ -\ 2\theta_4\partial_-
\ \ \ \ \ \ \  Q_3\ =\ \frac{\partial}{\partial\theta_3}
\end{array}}
$$
We remind that following \cite{we} we consider the reduction from
5-dimensional space to 2-dimensions, putting $\partial_1\ =\
\partial_2\ =\ \partial_3\ =\ 0$ and $\partial_4\ =\ \partial_+$,\  $\partial_5\ =\ \partial_-$.
We are going to concentrate our consideration on the first
non-trivial generator $Q_1$ which forms the closed algebra with the
generator $\partial_+$. The commutation relations are:
$$
\{\ Q_1\ ,\ Q_1\ \}\ =\ -4\partial_+
$$
\begin{equation}\label{algebra of Q1}
[\ Q_1\ ,\ \partial_+\ ]\ =\ 0
\end{equation}
$$
[\ \partial_+\ ,\ \partial_+\ ]\ =\ 0
$$
The idea is to add the sources $\varepsilon$ and $\eta$ for the
generators $Q_1$ and $\partial_+$ to the fundamental action
(\ref{Fund}) to form the BV action:
\begin{equation}\label{BV-SUSY action for BF}
S^{\scriptscriptstyle SUSY}\ =\ \int\ Tr\Big( <\EuScript{P},\ Q_{B}
\EuScript{A}>\ +\ g<\EuScript{P},\ \EuScript{A}^2>\ +\ \varepsilon<
\EuScript{P} ,\ Q_1 \EuScript{A}>\ +\ \eta<\EuScript{P},\
\partial_+\EuScript{A}>\ -\ 2\varepsilon^2\eta^{\ast}\ \Big)
\end{equation}
We remind that the field $\EuScript{A}$  is a generic superfield
build out of $\lambda^\alpha$, $\theta^\alpha$ and component fields.
$\EuScript{P}$ is a generic element of the space dual to the space
of superfields (dual superfield). The component fields of
$\EuScript{A}$  and $\EuScript{P}$ are different. Canonical pairing
$<\ ,\ >$ is defined as $< \underline{e}^a\ ,\ e_b\ >\ =\
\delta^a_b$. Here $e_a$ is a basis in the space of $\lambda$ and
$\theta$ and $\underline{e}^a$ is dual basis in the dual space.

By direct substitution one can check that this action satisfies
classical BV equation $$\int\ Tr\  \Bigg(\
\frac{\delta_{\scriptscriptstyle L} S}{\delta
\EuScript{A}}^{\!\scriptscriptstyle
SUSY}\!\frac{\delta_{\scriptscriptstyle R} S}{\delta
\EuScript{P}}^{\!\scriptscriptstyle SUSY}\ +\
\frac{\delta_{\scriptscriptstyle L} S}{\delta
\varepsilon}^{\!\scriptscriptstyle
SUSY}\!\frac{\delta_{\scriptscriptstyle R} S}{\delta
\varepsilon^\ast}^{\!\scriptscriptstyle SUSY}\ +\
\frac{\delta_{\scriptscriptstyle L} S}{\delta
\eta}^{\!\scriptscriptstyle SUSY}\!\frac{\delta_{\scriptscriptstyle
R} S}{\delta \eta^\ast}^{\!\scriptscriptstyle SUSY}\ \Bigg)\ =\ 0$$
The action does not depend on $\varepsilon^\ast$, hence the second
term in the Master Equation is automatically zero. Action
(\ref{BV-SUSY action for BF}) satisfies this equation under the
following conditions:
\begin{enumerate}
\item Operator $Q_B$ is nilpotent $Q_B^2\ =\ 0$.
\item Generators $Q_1$ and $\partial_+$ satisfy the commutation
relations (\ref{algebra of Q1}).
\item Operator $Q_B$ commutes with the generators as:
$\{Q_B,\ Q_1\}\ =\ 0$ and $[Q_B, \partial_+]\ =\ 0$.
\item Operators $Q_B$, $Q_1$ and $\partial_+$ differentiate multiplication of superfields $\EuScript{A}$, i.e. satisfy Leibnitz
identity.
\item The field $\varepsilon$ is even, the field $\eta$ is odd,
$\EuScript{A}$ and $\EuScript{P}$ are odd and even superfields
respectively.
\item The fields $\varepsilon$ and $\eta$ (as well as $\varepsilon^*$ and $\eta^\ast$)
are ghosts for the global symmetry, hence do not depend on
space-time coordinates.
\end{enumerate}
We are going to integrate out all the fields in the action
(\ref{BV-SUSY action for BF}) from the complement to the space of
cohomologies  $\mathcal{H}(Q)$.  Here operator $Q\ =\
\lambda_\alpha\frac{\partial}{\partial\theta_\alpha}$ is the first
term in (\ref{QBerkovits}).  These cohomologies were calculated in
the paper \cite{we} using the tower of fundamental relations, see
also \cite{Grassi} for the same calculation via localization
technique. Thus we are going to decompose the fields
\begin{eqnarray}
\EuScript{A}\ =\ \mathsf{A}\ +\ a\nonumber\\
\EuScript{P}\ =\ \mathsf{P}\ +\ p\nonumber
\end{eqnarray}
onto the superfield $\mathsf{A}$ of cohomologies of $\mathcal{H}(Q)$
and the superfield $a$ on the complement to $\mathcal{H}(Q)$. The
same decomposition on $\mathsf{P}$ and $p$ is done in the dual
space. The idea is to integrate over $a$ and $p$ to obtain effective
action on the component fields of $\mathsf{A}$ and $\mathsf{P}$.

 The sources $\varepsilon$ and $\eta$ for the supersymmetry and
translations will allow to control the supersymmetric properties of
the effective action. The next subsection contains the calculation
of effective action using the Feynman diagram technique established
in \cite{we}.

\newpage
\subsection{Calculation of Effective action}
\label{Sub-sec Calc  S^eff}
\begin{wraptable}[32]{l}{200pt}
$$
\begin{array}{|c|c|c|}
\hline
$Polarization$  & $Field$ & $Antifield$ \\
\hline
 & &\\
1  & c & \widetilde{c} \\
 & &\\
 \hline
 & &\\
\lambda_1\theta_2\ +\ \lambda_2\theta_1  & \gamma_+ & \widetilde{\gamma}_+ \\
\lambda_2\theta_3\ +\ \lambda_3\theta_2  & \varphi & \widetilde{\varphi} \\
\lambda_3\theta_4\ +\ \lambda_4\theta_3  & \gamma_- & \widetilde{\gamma}_- \\
\lambda_1\theta_1  & A_+ & \widetilde{A}_+ \\
\lambda_4\theta_4   & A_- &  \widetilde{A}_- \\
 & &\\
\hline
 & &\\
\lambda_1\theta_1\theta_2  & \psi_+ & \widetilde{\psi}_+ \\
\lambda_4\theta_4\theta_3  & \psi_- &  \widetilde{\psi}_-\\
 \lambda_2\theta_3\theta_1  & \chi_+ & \widetilde{\chi}_+ \\
\lambda_3\theta_2\theta_4  & \chi_- & \widetilde{\chi}_- \\
 & &\\
\hline
 & &\\
\lambda_1\lambda_4\theta_1\theta_4  & \varphi_1 & \widetilde{\varphi}_1 \\
\lambda_1\lambda_4\theta_4\theta_2\ +\ \lambda_2\lambda_4\theta_4\theta_1  & \varphi_2 & \widetilde{\varphi}_2 \\
\lambda_1\lambda_3\theta_1\theta_4\ +\ \lambda_1\lambda_4\theta_1\theta_3  & \varphi_3 & \widetilde{\varphi}_3 \\
 & &\\
\hline
 & &\\
\lambda_1\lambda_4\theta_1\theta_4\theta_3  & \varphi_4 & \widetilde{\varphi}_4 \\
\lambda_1\lambda_4\theta_1\theta_2\theta_4  & \varphi_5 & \widetilde{\varphi}_5 \\
\lambda_1\lambda_3\theta_1\theta_2\theta_4\ +\ \lambda_1\lambda_4\theta_1\theta_2\theta_3  & \varphi_6 & \widetilde{\varphi}_6 \\
\lambda_2\lambda_4\theta_1\theta_4\theta_3\ +\ \lambda_1\lambda_4\theta_2\theta_4\theta_3  & \varphi_7 & \widetilde{\varphi}_7 \\
 & &\\
\hline
 & &\\
\lambda_1\lambda_4\theta_1\theta_2\theta_3\theta_4  & \varphi_8 & \widetilde{\varphi}_8 \\
 & &\\
\hline
\end{array}
$$
\end{wraptable}
\

For the calculations in this section we use the notations of section
6 of \cite{we}. The physical degrees of freedom -- representatives
of cohomologies of operator $Q$, are presented in the table. The
first column gives polarizations for
 the component fields, the
second one - notations for the component fields of $\mathsf{A}$, the
last one -  notations for the component fields of $\mathsf{P}$. To
compute effective action one has to sum up all connected tree
diagrams with the external legs being the component fields of
$\mathsf{A}$ (input lines in the diagram) and $\mathsf{P}$ (the
output line). As it was explained in \cite{we} each diagram can have
only one output line.

  Conducting this calculation one has to remember that the diagrams
having the propagator (wavy line), like the diagrams 4,5,6,7 in the
figure \ref{figure linear}, should be added with the relative
\textit{minus} sign to the diagrams without propagator, like the
diagrams 1,2,3 in the figure \ref{figure linear}. The simplest
argument for this can be given in Minkowski space. Each operator
insertion, like $\Phi$, $\varepsilon Q_1$ or $\eta\partial_+$,
contributes a factor of $i$ (complex unity), coming from the
exponent, the propagator (wavy line) also contributes a factor of
$i$. Hence the diagrams 1,2,3 are proportional to $i$, while the
diagrams 4,5,6,7 are proportional to $-i$ (there are two operators
and one propagator). That is why to find the result for the
effective action one has to sum all the diagrams without propagator
and subtract all the diagrams with one propagator. It is
straightforward to demonstrate that it is impossible to draw the
diagrams with more than one propagator by calculating the degree in
$\lambda$ and $\theta$ in the final expression before taking the
projection onto cohomologies.


\subsubsection{Linear level}
\label{sub-sub-sec. calc S^eff, liniar}

\begin{figure}[h]
\rightline{\includegraphics[width=160mm]{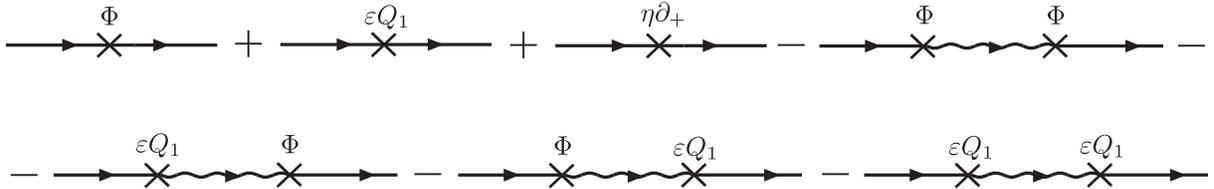}}
\caption{{\footnotesize Linear level.}} \label{figure linear}
\end{figure}

Firstly, consider the linear problem (gauge coupling constant $g =
0$). The list of the diagrams giving nonzero result is presented in
the fig. \ref{figure linear}.

\

A lot of diagrams are absent in this figure. To prove this fact one
has to count the degrees of lambda and theta. For example, the
diagrams with three insertions of operator $\Phi$  are absent. This
is true because each insertion of operator $\Phi$ gives
multiplication by $\theta
 \lambda$, and these diagrams should have at least 2 propagators, each carrying
degree  $\theta / \lambda$. Thus these diagrams should change the
degree in $\lambda$  and $\theta$ by: $(\frac{\theta}{\lambda})^2
(\lambda\theta)^2\ =\ \theta^4$. However, there are no two
representatives of cohomologies, having equal number of $\lambda$
and difference $4$ in the degree of $\theta$.

Another example  of vanishing  sub-diagrams is presented in the fig.
\ref{figure forbidden  subdiagrams}. In the first fragment
propagator is applied to cohomology, this diagram is equal to zero.
Really,  propagator gives non-zero result only if it is applied to
$Q$-exact expression (see section 3 of \cite{we} for details).
\begin{figure}[h]
\centerline{\includegraphics[width=100mm]{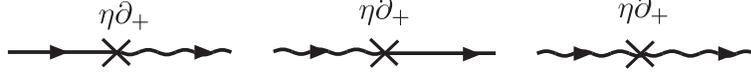}}
\caption{{\footnotesize Forbidden subdiagrams.}} \label{figure
forbidden  subdiagrams}
\end{figure}
The same is true for the second fragment because the image of the
propagator has zero projection onto cohomologies. In the last
fragment we meet square of propagator which is equal to zero.
 Returning to the diagrams in the fig.~\ref{figure linear} we recall that
the first and the last diagrams in the first line  does not contain
ghosts for SUSY or translations, hence they have been already
calculated in \cite{we}. Below we give a schematic illustration of
the procedure for calculation of other diagrams.

1.  The simplest one is with the insertion of $\eta \partial_+$
operator(the third diagram in the fig.~\ref{figure linear}). Each of
18 fields contributes to this diagram in  a trivial way: if input
line is the certain field then output line exactly projected to its
antifield.

\centerline{\includegraphics[width=80mm]{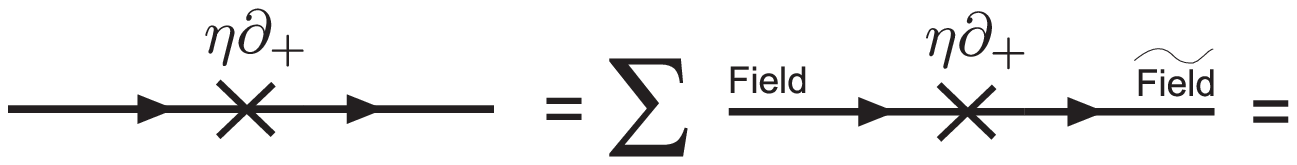}}
\vspace{-0.5cm}
 \be =\eta ( \widetilde{c} \partial_+ c + ... +
\widetilde{\varphi}_8
\partial_+ \varphi_8) \ee

2. Consider the diagrams with the insertion of operator $Q_1$ (the
second diagram in the fig.~\ref{figure linear}).  Among all 18
fields of the theory only 8 give contribution to this diagram. They
are: $\gamma_+, \ \chi_+,\ \psi_+,$ $\ \varphi,\ \varphi_2 ,\
\varphi_5,\ \varphi_7 ,\ \varphi_8$. Consider for instance
$\gamma_+$ as an input line: \be \gamma_+: \e Q_1(\lambda_1 \theta_2
+ \lambda_2 \theta_1)\gamma_+ = \varepsilon (\lambda_2 \gamma_+ - 2
\lambda_1 \theta_1\theta_2\
\partial_+\gamma_+) \xrightarrow{Projection} -2 \varepsilon
\,\widetilde{\psi}_+
\partial_+ \gamma_+\nn. \ee To project the result written on the l.h.s of the arrow one
should look at the 7$^{th}$ line of the table with the polarizations
(cohomologies). Conducting the same procedure, one can obtain the
contributions of all 8 fields to the effective action. The result is
given by: \be \Delta S_{Lin}^{(Q_1)} = \int Tr \Big( - 2 \varepsilon
\widetilde{\psi}_+
\partial_+ \gamma_+ \ + \ 2\varepsilon \widetilde{\varphi}_8
\partial_+ \varphi_7  +  2 \varepsilon
\widetilde{\varphi}_5
\partial_+ \varphi_2  + \frac1 2 \varepsilon \widetilde{\gamma}_+ \psi_+  - \frac1 2
\varepsilon \widetilde{\varphi} \chi_+\nn - \frac1 2\varepsilon
\widetilde{\varphi}_2 \varphi_5   - \frac1 2\varepsilon
\widetilde{\varphi}_7 \varphi_8 \Big) \ee

A little bit more difficult is to calculate the diagrams with the
propagator. The propagator $K$ is defined in the section 3 of
\cite{we}. Roughly speaking it acts as follows: $K$, acting on
representatives of cohomologies gives zero; being applied to exact
expressions propagator gives the pre-image ($K(Q\omega)\ =\ \omega$)
of operator $Q$; propagator, being applied to certain non-closed
expressions gives zero (see \cite{we} for details).

 3. The diagrams
containing $\Phi$ and $Q_1$ (the 5$^{th}$ and 6$^{th}$ diagrams in
the fig.~\ref{figure linear}). In case $\Phi Q_1$-diagrams ( $Q_1$
acts first), input lines are $ \gamma_+,\ \varphi_2,\ \varphi_7$.
Consider for example $\gamma_+$: \be \Phi K \varepsilon Q_1
(\lambda_1 \theta_2 + \lambda_2 \theta_1) \gamma_+ = \varepsilon
\Phi K (\lambda_2 \gamma_+ - 2 \lambda_1 \theta_1\theta_2\
\partial_+\gamma_+) = \varepsilon \Phi K \Big(Q (\theta_2\gamma_+) - 2 \lambda_1 \theta_1\theta_2\
 \partial_+ \gamma_+\Big)\nn ,\ee
 the second term in the  r.h.s. is proportional
 to the  cohomology and  propagator acts on it as zero.
\be \varepsilon \Phi K \Big(Q (\theta_2\gamma_+) - 2 \lambda_1
\theta_1\theta_2\
 \partial_+ \gamma_+\Big) =  \varepsilon \Phi \theta_2 \gamma_+ =  2\varepsilon \lambda_1 \theta_1\theta_2 \partial_+ \gamma_+ +
 2\varepsilon \lambda_4\theta_4\theta_2 \partial_- \gamma_+
 \ee%
In this simple example one can  see the key feature of this
calculation. The first term in the r.h.s. has the polarization of
$\widetilde{\psi}_+$ (see line 7 in the table) hence one can project
this term  to $2 \varepsilon \,\widetilde{\psi}_+
\partial_+ \gamma_+\nn $. The second term is nonclosed, but it was mentioned before
each expression has a chance to sum up with  similar term from
another diagram to form  a closed result. This closed result can
have a non-trivial  projection onto cohomologies. The result for the
remaining two input lines and $\gamma_+$ is the following \be \Delta
S_{Lin}^{(\Phi Q_1)} = \int Tr \Big(
  2\varepsilon \widetilde{\varphi}_8
\partial_+ \varphi_7 - 2 \varepsilon \widetilde{\psi}_+
\partial_+ \gamma_+ +   2 \varepsilon
\widetilde{\varphi}_5
\partial_+ \varphi_2 \Big).
\ee Finally, in case $ Q_1 \Phi$ ($\Phi$ acts first) there are no
proper input fields resulting in the closed expression.

4. Diagrams containing $Q_1 Q_1$. These  diagrams  are the most
interesting in the sense, that they give terms proportional to
$\varepsilon^2$. From the section 2 we know how to interpret this
terms. Possible input lines for this diagram are: $A_+,\ \varphi_1,\
\varphi_3,\ \varphi_4,\ \varphi_6$. The resulting contribution to
the effective action is

\be%
 \Delta S_{Lin}^{(Q_1 Q_1)} = \int Tr \Big( -
\varepsilon^2\widetilde{c}A_+ - \varepsilon^2 \widetilde{A}_-
\varphi_1 - \varepsilon^2\widetilde{\gamma}_- \varphi_3 -
\varepsilon^2 \widetilde{\psi}_- \varphi_4 - 2 \varepsilon^2
\widetilde{\chi}_-\varphi_6 \Big).\nn
 \ee%

5. After all one should examine  all nonclosed results coming  from
all the diagrams. There are  six such terms. Corresponding in-lines
are $\gamma_+,\ \gamma_-,\ A_+,\ A_-,\ \varphi$. Among these six
terms  only the field  $\varphi$ is summed up into a non-zero
contribution into effective action. One nonclosed part comes from
$Q_1 \Phi$ diagram, another part comes from nonpropagating diagram
with $Q_1$ inserted.
\bee%
&\varepsilon Q_1 \varphi  = -2 \varepsilon ( \lambda_2
\theta_1\theta_3 + \lambda_3 \theta_1\theta_2 )\partial_+ \varphi
\nn \\%
&\varepsilon Q_1 K  \Phi\, \varphi  =  2 \varepsilon \lambda_1
\theta_2\theta_3 \partial_+ \varphi
\nn \\%
&\varepsilon (Q_1 - Q_1  K \Phi)\varphi  = - 2 \varepsilon
(\lambda_2 \theta_1\theta_3 + \lambda_1 \theta_2\theta_3 + \lambda_3
\theta_1\theta_2) \partial_+ \varphi = 2\varepsilon \Big( 2
\lambda_2 \theta_3\theta_1 - Q( \theta_1 \theta_2\theta_3) \Big)
\xrightarrow{Proj.} 4 \varepsilon \widetilde{\chi}_+ \partial_+
\varphi,
\eee%
since projection to cohomologies annihilates exact expressions.

Now we are ready to write down the whole linear effective action
with  the sources $\varepsilon$ and $\eta$.
\be%
\label{Action linear effective with sources}
\begin{split}
S^{eff}_{lin}\ = &\ \int Tr\bigg( \ \ \widetilde{\varphi}_1\Big(\
2(\partial_+A_- -
\partial_-A_+)  \ \Big)\  +\ \widetilde{\varphi}_2\Big(\ \ 2\partial_-\gamma_+ - \frac1 2 \e \,  \varphi_5\ \Big)
 \ +\ \widetilde{\varphi}_3\Big(\ 2\partial_+\gamma_-\ \Big)\ +\ \widetilde{\varphi}_4\Big(\
2\partial_+\psi_-\ \Big)\ +
\\ &+ \ \widetilde{\varphi}_5\Big(\
2\partial_-\psi_+ + 4 \e \, \partial_+ \varphi_2\ \Big)\ +\
\widetilde{\varphi}_6\Big(\
\partial_+\chi_-\ \Big)\  +\ \widetilde{\varphi}_7\Big(\
\partial_-\chi_+\ - \frac1 2 \e \, \varphi_8  \Big)\ + \
\widetilde{\varphi}_8\Big( + 8\partial_+\partial_-\varphi\ + 4 \e \,
\partial_+ \varphi_7  \Big) +
\\&+ \ \widetilde{\gamma
}_+ \Big(\ \frac1 2 \e \, \psi_+ \ \Big)\  +\ \widetilde{\gamma }_-
\Big(\ - \e^2  \varphi_3 \ \Big)\ + \ \widetilde{\varphi } \Big(\
-\frac1 2 \e \,\chi_+ \ \Big)\ + \ \widetilde{A}_+ \Big(\  2
\partial_+ c\ \Big)\ + \ \widetilde{A }_- \Big(- \e^2
\varphi_1 + 2 \partial_- c \ \Big)+
\\& + \ \widetilde{\psi
}_+ \Big( - \ 4 \e \, \partial_+ \gamma_+ \ \Big)  + \
\widetilde{\psi }_- \Big(- \e^2 \varphi_4 \ \Big)\ + \
\widetilde{\chi }_+ \Big(\ 4 \e \, \partial_+ \varphi \ \Big)\ + \
\widetilde{\chi }_- \Big(-2 \e^2 \varphi_6 \ \Big)  + \ \widetilde{c
} \Big(- \e^2 A_+\ \Big) \ +
\\& +\  \eta \  \Big( \widetilde{c} \, \partial_+ c \ +\  ...\  + \
\widetilde{\ \varphi}_8 \partial_+ \varphi_8 \Big) \ - \  2 \eta^*
\, \e^2\ \bigg)
\end{split}
\ee%
\subsubsection{Nonlinear level}
\label{sub-sub-sec. calc S^eff, nonliniar} Fortunately there is
quite small number of additional diagrams arising after switching on
the interaction. They are depicted in fig. \ref{figure nonlinear}.
\begin{figure}[h]
\centerline{\includegraphics[width=120mm]{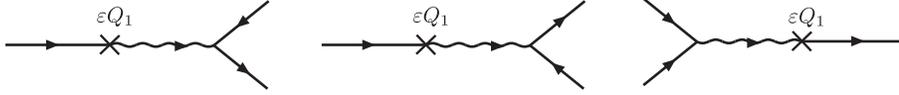}}
\caption{{\footnotesize Nonlinear diagrams.}} \label{figure
nonlinear}
\end{figure}
All calculations are  completely analogous to linear case.

One can straightforwardly check   that only <<$\varepsilon g$>>
order survives. As in the previous case two pairs of nonclosed
constructions find each other and result into
\be%
\Delta S_{Nonlin}^1 = \int Tr \  \Big (2 \varepsilon
 g \widetilde{\chi}_+ [A_+,\varphi] \ - 4  \varepsilon
 g \widetilde{\chi}_-
[\gamma_+,\gamma_-]\ \Big)
\ee%
other terms are "purely"  projected, i.e. each diagram gives closed
result and can be projected separately without summing up with
another diagram.
\be%
\begin{split}
\Delta S_{Nonlin}^2 = \int Tr \  \Big ( - 2 \varepsilon  g \
\widetilde{\psi}_+[A_+,\gamma_+] \ - \ 2\varepsilon  g \
\widetilde{\varphi}_5 [\varphi_1,\gamma_+] \ -\ 2\varepsilon  g \
\widetilde{\varphi}_6 [\varphi_3,\gamma_+] \ +\ 2\varepsilon  g \
\widetilde{\varphi}_8 [\varphi_4,\gamma_+] \ -\\- 2\varepsilon  g \
\widetilde{\varphi}_5 [\varphi_2,A_+]  \ - \  2\varepsilon  g \
\widetilde{\varphi}_8 [\varphi_7,A_+]\
 + \ \frac1 2 \varepsilon  g \
\widetilde{\varphi}_8 \{\varphi_1,\chi_+\} \ -\ \frac1 2 \varepsilon
g \ \widetilde{\varphi}_7 [\varphi,\varphi_1]\  \ \Big)
\end{split}
\ee%

%

 Finally we collect together the results of the work \cite{we} and additional
 terms with sources for SUSY. Effective lagrangian is given by
\be%
\label{Final action for 2d model}
\begin{split}
L^{eff}      &=  \bigg[ \Phi F_{+-} + \varphi_8 \{D_+,D_- \} \varphi
- g
    \varphi_8\{\psi_+,\psi_-\}
    + 2 g \bar{\chi}_- [\gamma_-,\psi_+]  +  2 g \bar{\chi}_+ [\gamma_+,\psi_-]
      + \beta_+ D_- \gamma_+ + \beta_- D_+ \gamma_-
      \\&+ \bar{\psi}_-
    D_+ \psi_-
    + \bar{\psi}_+ D_- \psi_+ +  \bar{\chi}_- D_+ \chi_- + \bar{\chi}_+ D_- \chi_+ \bigg]%
     +  \bigg[2A_+^* \partial_+c +2  A_-^* \partial_-c
    + g \Big( \  c^* c c
    + ... +\varphi^*_8\{\varphi_8,c\}  \ \Big)\bigg] \\&
    +  \eta  \ \Big[
    c^* \, \partial_+ c \ +\  ...\  + \ \
    \varphi_8^* \partial_+ \varphi_8 \Big]
     +  \varepsilon\bigg[   2 \chi_+^* D_+ \varphi - \varphi_8 D_+
    \bar{\chi}_+^*
    - 2 \psi_+^* D_+ \gamma_+ -  2\bar{\psi}_- D_+ \beta_+^* +\frac1 2 \beta_+ \bar{\psi}_+^*
    \\&+ \chi_+ \varphi_8^* + \frac1 2 \gamma_+^*\psi_+ - \frac1 2 \varphi^* \chi_+
     - g\Big( \bar{\chi}_+ [\varphi, \Phi^*] - 2 \bar{\psi}_+ [\Phi^*, \gamma_+] -
    4\bar{\chi}_-[\beta_-^*,\gamma_+] + \frac1 2 \varphi_8\{\Phi^*,
    \chi_+\} - 2 \varphi_8 [\gamma_+, \bar{\psi}_-^*]  \\& + 4 \psi_+^*[\gamma_+,\gamma_-]
    \Big) \bigg]
     - \ 2\eta^*\varepsilon^2 + \varepsilon^2  \bigg[   \gamma_-^* \beta_-^* + A_-^* \Phi^*
     +
     \psi_-^* \bar{\psi}_-^* + \chi_-^* \bar{\chi}_-^* - c^* A_+ \bigg].
\end{split}
\ee Here we turn to the physical notations as it was in \cite{we}.
Namely, $$D_+ = 2\p_+ +g[A_+,\cdot],\ \ \ \  D_- = 2\p_-
+g[A_-.\cdot], $$ $$F_{+-} = 2(\p_-A_+ -\p_+A_-) + g[A_+,A_-]$$ and
fields $$c, A_{\pm}, ... , \chi_{\pm} \rightarrow c, A_{\pm}, ... , \chi_{\pm}$$
$$\widetilde{c}, \widetilde{A}_{\pm}, ... , \widetilde{\chi}_{\pm}
\rightarrow c^*, A_{\pm}^*, ... , \chi_{\pm}^* $$
$$
\widetilde{\varphi}_1,\widetilde{\varphi}_2, ... ,
\widetilde{\varphi}_7, \widetilde{\varphi}_8  \rightarrow \Phi,
\b_+, \b_-,\bar{\psi}_-,\bar\psi_+ , 2\bar\chi_-, 2\bar\chi_+,
\varphi_8$$
$$\varphi_1,
\varphi_2,...,\varphi_7, \varphi_8 \rightarrow -\Phi^*,- \b_+^*,-
\b_-^*,-\bar{\psi}_-^*,-\bar\psi_+ ^*,- \frac12\bar\chi_-^*,-
\frac12\bar\chi_+^*, -\varphi_8^* $$
One can see that the terms discussed in sections 2 and 3 appear in the last  brackets.
 According to the discussions above, these terms are responsible for the descent of the off-shell SUSY
 invariance of (\ref{Fund}) down to on-shell SUSY invariance  of (\ref{Final action for 2d model}).

\section{Berkovits' 10-d Super Yang-Mills}
\label{section SYM} In this section we apply the ideas developed
earlier in this paper to the 10-d Super Yang-Mills theory
\cite{Berkovits}, \cite{ced}. This theory is more interesting from
the physical point of view than the model considered in the previous
section. However, the off-shell description\footnote{The problem of
off-shell formulation in the context of harmonic superspace was
studied in \cite{harmonic}.} of this model is more complicated
because of the necessity to make $Z_2$ projection on the space of
fields.

  According to the ideology from the previous sections one should
calculate effective action for the theory
\be%
\label{10d  SYM} S^{SUSY} = \int\ Tr\Big( <\EuScript{P},\  Q_{B}
\EuScript{A}> \ +\  g <\EuScript{P},\  \EuScript{A}^2\!>\ + \
<\EuScript{P},\ \e^\a  Q^s_\a \EuScript{A}>\ + \ \!<\EuScript{P},\
\eta^\mu P^{s}_\mu \EuScript{A}>\  -\ \ \eta_\mu^*(\e\g^\mu\e) \Big)
\ee%
on the cohomologies of operator $Q\ =\
\lambda^\alpha\frac{\partial}{\partial\theta^\alpha}$. Here we use
the following notations
\begin{equation}
 Q_B \ =\  \lambda^\alpha \frac{\partial}{\partial
\theta^\alpha}+ \frac12 \theta^\alpha \frac{\partial f^\mu}{\partial
\lambda^\alpha} \partial_\mu,\ \ \ \  \e^\a Q^s_\a \ =\  \e^\a
\frac{\p}{\p \t^\a} - (\e\g^\mu\t) \frac{\p}{\p x^\mu} , \ \ \
P^s_\mu \ =\ \frac{\p}{\p x^\mu}
\end{equation}
This calculation is done through the summation of all possible
Feynman diagrams according to the standard rules discussed in
section 4 (see also \cite{we}). The degrees of freedom
(representatives of $Q$-cohomologies) for this model are given by
\be%
\label{Spectrum of YM(table)}%
\begin{array}{cc|ccc}
\hline
\text{ Polarization} &           \mathsf{A}   &  & \text{ Dual polarization}        &\mathsf{P}\\
\hline 1 & c & & \underline{1} &\widetilde{c}\\
 (\lambda\gamma^{\mu}\theta)&A_{\mu}&&\underline{(\lambda\gamma^{\mu}\theta)}&\widetilde A_{\mu}\\
 (\lambda\gamma^{\mu}\theta)(\theta\gamma^{\mu})_{\alpha}&\psi^{\alpha}&&\underline{ (\lambda\gamma^{\mu}\theta)(\theta\gamma^{\mu})_{\alpha}}
 &\widetilde{\psi}^{\alpha}
 \\
-16
(\lambda\gamma^{\mu}\theta)(\lambda\gamma^{\nu}\theta)(\theta\gamma^{\mu\nu})^{\alpha}&
\psi^*_{\alpha} && -16\underline{
(\lambda\gamma^{\mu}\theta)(\lambda\gamma^{\nu}\theta)(\theta\gamma^{\mu\nu})^{\alpha}} &\widetilde{\psi}^*_\alpha \\
10
(\lambda\gamma^{\mu}\theta)(\lambda\gamma^{\nu}\theta)(\theta\gamma_{\mu\nu\rho}\theta)&
A_\rho^*&&10\underline{
(\lambda\gamma^{\mu}\theta)(\lambda\gamma^{\nu}\theta)(\theta\gamma_{\mu\nu\rho}\theta)}
&\widetilde{A}_\rho^*\\
(\lambda\gamma^{\mu}\theta)(\lambda\gamma^{\nu}\theta)(\lambda\gamma^{\rho}\theta)
  (\theta\gamma_{\mu\nu\rho}\theta)&c^*&&\underline{(\lambda\gamma^{\mu}\theta)(\lambda\gamma^{\nu}\theta)(\lambda\gamma^{\rho}\theta)
  (\theta\gamma_{\mu\nu\rho}\theta)}&\widetilde{c}^*\\

\end{array}
\ee%
The first and the third columns contain the polarizations
(representatives of cohomologies) for the fields and antifields
respectively, the second column gives the component fields of the
superfield $\mathsf{A}$ and the fourth one the component fields of
the superfield $\mathsf{P}$. The component fields in the fourth
column are BV antifields to the component fields in the second
column. For example $\widetilde{c}$ is BV antifield to $c$. The same
is true for other fields. Let $\{e_B \}$ denote the basis in the
space of functions of $\l$ and $\t$. Let $\{\underline{e}^A \}$
denote the dual basis in the dual space. There is a canonical
pairing among them, that we denote as $<\ ,\ >$: $<\underline{e}^A,
e_B> = \d^A_B$. For example
$$
<\underline{1}, 1 > = 1,
\ \ \ \ \ \ \ \ \
<\underline{1}, (\l\g^\mu\t) > = 0
$$

\subsection{Doubling}
From the table (\ref{Spectrum of YM(table)}) we see that the number
of fields and anti-fields in the theory is twice the number of the
fields we expect to have in SYM.  The second column contains all the
fields needed for BV version of SYM (this is true if one can think
about the fields with the star-sign as antifields for the
corresponding fields). In addition to them there are their BV
antifields marked by tilde-sign (the content of the fourth column).
At the present moment BV bracket is simply the canonical pairing,
between $\mathsf{P}$ and $\mathsf{A}$.

We will call this theory, effective for (\ref{10d  SYM}), the {\bf
pre-SYM}. This funny name is due to the fact that the pre-theory and
the theory (SYM) are related by the simple transformation which will
be discussed in  section \ref{Z2 projection subsection} below.
Roughly speaking, in order to get SYM itself one should identify
some fields and antifields of  the effective pre-theory as it is
shown in the table (\ref{Spectrum of YM(table) 2}).

\subsection{Technical Subtleties in the Calculation of Pre-SYM Action}
In the calculation we follow the standard technique. However it is
technically complicated to project on cohomologies of $Q$. Instead
we implement  the following procedure \cite{Mafra-Berk}.

Consider the space of functions of the 3-rd power in $\lambda$ and
the 5-th power in $\theta$. Consider the subspace of this space
$V_1$ generated by two elements
$$
(\lambda\gamma^{\mu}\theta)(\lambda\gamma^{\nu}\theta)(\lambda\gamma^{\rho}\theta)(\theta\gamma_{abc}\theta)
$$
$$
(\lambda\gamma^{\mu\nu\rho}\theta)(\lambda\gamma_{p}\theta)(\lambda\gamma_{q}
\theta)(\theta\gamma_{abc}\theta)
$$
This space $V_1$ can be decomposed into the sum of irreducible
representations. The only cohomology in this space is $h_{3,5}\ =\
(\lambda\gamma^{\mu}\theta)(\lambda\gamma^{\nu}\theta)(\lambda\gamma^{\rho}\theta)(\theta\gamma_{\mu\nu\rho}\theta)$,
which is a  scalar. Consider linear functional $\bl\ \  \br$ on the
space $V_1$ such that it maps cohomology $h_{3,5}$  to $1$ and
non-trivial representations to zero. Namely \cite{Mafra-Berk}
\bee%
 \ll\!\!(\lambda\gamma^{\mu}\theta)(\lambda\gamma^{\nu}\theta)(\lambda\gamma^{\rho}\theta)(\theta\gamma_{abc}\theta)\!\!\gg
 = \frac1{120} \delta^{\mu\nu\rho}_{abc}\\
  \ll\!\!(\lambda\gamma^{\mu\nu\rho}\theta)(\lambda\gamma_{p}\theta)(\lambda\gamma_{q} \theta)(\theta\gamma_{abc}\theta)\!\!\gg
 = \frac1{70} \delta^{[\mu}_{[p}\eta_{q][a}^{}\delta^\nu_b \delta^{\rho
 ]}_{c]}\nonumber
\eee%
The coefficient is restored from the condition that the cohomology
$h_{3,5}$ is mapped to unity.

 In the computations that we perform we replace projection to
cohomologies by the following procedure. For each representative $h_a$
from the table (\ref{Spectrum of YM(table)}) define a complementary
representative $d_b$ such that\footnote{Since the bracket $\bl\ \ \
\br$ maps all functions having the degree in $\l$ and $\t$ different
from 3 and 5 to zero, such complementary representative is unique and
completes the degree in $\l$ and $\t$ to $3,5$.}
\begin{equation}%
\label{Z2 duality on representatives}%
\bl h_a\cdot d_b\br \ =\ \d_{ab}
\end{equation}
Here the product is induced by the multiplication of functions of
$\lambda$ and $\theta$. To project some expression $\Omega$ onto the
representative $h$ one should instead calculate the product with the
complementary representative $d$
\be%
\label{rule of projection}%
\Omega \big|_{h_a} =\  (-1)^{\#\!d_a +1} \bl d_a\cdot\Omega \br
\ee%
Here $\#\!d_a $ denotes the parity of the representative $d_a$. It
happens that with our choice of representatives and $K$ the result
before projection is always in the space $V_1$ for all the Feynman
diagrams.

 Using this prescription one can calculate the effective action for
the theory (\ref{10d  SYM}). The result is given by
\be%
\begin{split}
\label{pre effective action}%
L_{pre}^{eff} =  - \frac1{360} \widetilde{A}^*_\mu D_\nu F_{\mu\nu}
+ \frac1{160} \psi\gamma^\mu D_\mu \widetilde{\psi}^*  + \frac
g{160} \widetilde{A}_\mu^*(\psi\gamma^\mu \psi)   +
\widetilde{A}_\mu D_\mu c +  A^*_\mu D_\mu \widetilde{c}^* -  g
A^*_\mu [ \widetilde{A}_\mu^*,c]
  +
  \\ + g \big( \widetilde{c}cc + \widetilde{c}^*[c^*,c]  +
  [\widetilde{\psi}^*,\psi^*]c +
[\widetilde{\psi},\psi]c + \widetilde{c}^*[\psi^*,\psi]\big) +
\\
 +\frac3{2} (\varepsilon\gamma^\mu\psi) \widetilde{A}_\mu
 +\frac3{2} (\varepsilon\gamma^\mu\widetilde{\psi}^*) A_\mu^*
  + \frac2{3}
(\varepsilon \gamma^{\mu\nu}\widetilde{\psi})D_{\mu}A_\nu +\frac2{3}
(\varepsilon \gamma^{\mu\nu}\psi^*)D_{\mu}\widetilde{A}^*_\nu -\frac
g{3} (\varepsilon \gamma^{\mu\nu}\widetilde{\psi})[A_{\mu},A_\nu]+
\\ + \eta^\mu\Big[\widetilde{c}\p_\mu c - c^*\p_\mu \widetilde{c}^* +  \widetilde{A}_\nu\p_\mu A_\nu -
 A_\nu^*\p_\mu \widetilde{A}^*_\nu  +
(\widetilde{\psi}\p_\mu\psi) - (\psi^*\p_\mu\widetilde{\psi}^*)\Big]
-\eta^*_\mu (\e\g^\mu\e)-
\\
- 80
(\varepsilon\gamma^\mu\varepsilon)(\widetilde{\psi}\gamma_\mu\psi^*)
+ 160 (\varepsilon\widetilde{\psi})(\varepsilon\psi^*)
-c^*(\varepsilon\gamma^\mu\varepsilon)\widetilde{A}^*_\mu
-\widetilde{c}(\varepsilon\gamma^\mu\varepsilon)A_\mu
\end{split}
\ee%
The details of this calculation can be found in the appendix B.

\noindent  Here $D_\mu = \p_\mu + g[A_\mu, \cdot]$, $F_{\mu\nu} =
\p_\mu A_\nu - \p_\nu  A_\mu + g [A_\mu,A_\nu]$.

We would like to emphasize that construction \cite{Mafra-Berk} with
the bracket $\bl \ \ \  \br$ is nothing but only a technical
simplification in the way to project onto cohomologies.

\subsection{$Z_2$ Duality of Feynman Diagrams}
\begin{figure}[h]
 \centerline{\includegraphics[width=140mm]{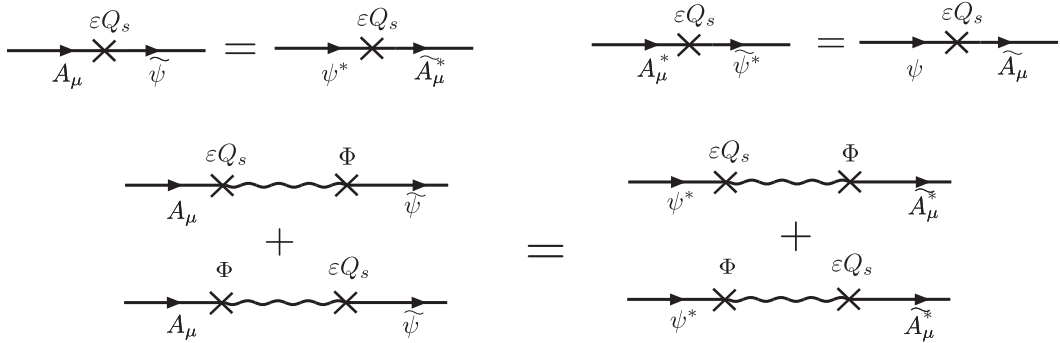}}
\caption{{\footnotesize An example of $Z_2$ duality. Explicit calculations are presented in the appendix B.}}
\label{figure z2}
\end{figure}

The correspondence between the representatives (\ref{Z2 duality on
representatives}) defines the $Z_2$ symmetry on cohomologies (the
$Z_2$ symmetric representative is the one which completes the given
one to non-vanishing value of the bracket $\bl\ \ \  \br$).  Though
this $Z_2$ symmetry on representatives is explicit (as is obvious
from the table \ref{Spectrum of YM(table)} and was discussed in
\cite{Gorod-Rud}) it is unclear why it should be inherited by
diagrams calculations. As a kind of {\it experimental} evidence of
this fact  below (fig.\ref{figure z2}) we present the results for
several diagrams arising in the calculation of effective action at
the quadratic level. From this figure it is clear that the vertices
of the effective theory are symmetric w.r.t. the discussed $Z_2$
duality.

One comment is in order here. Since we are interested in the action of SYM
(not pre-SYM) we will finally identify  the component fields of $\mathsf{A}$ and $\mathsf{P}$ according to
(\ref{Spectrum of YM(table) 2}). Mnemonic rule is the following: tilde and star is the same and tilde annihilates star.
 To understand the equalities in the figure \ref{figure z2} correctly one should make these identifications (\ref{Spectrum of YM(table) 2}).
The same is true for figure \ref{figure duality rule}.

Note that these equalities express a non trivial statement, because
incoming and the out-coming lines of the diagrams are completely
different. All the operators are acting  on the incoming lines and
the result is projected on the out-coming line. Moreover, among
these operators
 there are derivatives  w.r.t $\t$ (in $\e^\a Q_\a^s$). This duality states that there is, in a sense, a symmetry between the incoming
 and out-coming arrows. This symmetry results in the symmetry of the vertices of the effective action after the
 identification (\ref{Spectrum of YM(table) 2}).

\textbf{Duality rule.} For generic subset of diagrams having certain
external legs one can exchange  the  out-coming line with any of the
in-lines simultaneously changing the star and tilde signs and shifting external legs in the  cyclic way as it
 is shown  in the  fig. \ref{figure duality rule}. The result does not change.
It is important to remember that before application of this duality
one should sum up all the diagrams having certain external legs
(like in case of the last diagrams in the figure \ref{figure z2}).
\begin{figure}[h]
\begin{center}
\includegraphics[width=45mm]{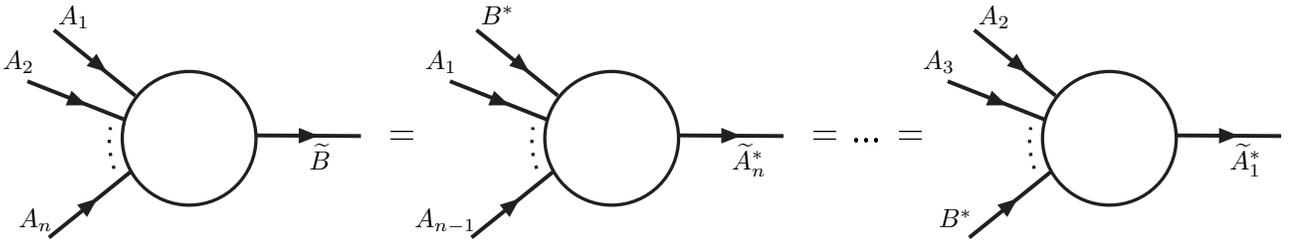}
\put(-115,82){$A_1$}\put(-136, 65){$A_2$}
\put(-130,5){$A_n$}\put(-20, 27){$\widetilde{B}$}  \put( 10,37){\bf
\large = } \hspace{1cm}
\includegraphics[width=45mm]{figure_z-2.eps}
\put(-115,82){$B^*$}\put(-136, 65){$A_1$}
\put(-140,5){$A_{n-1}$}\put(-20, 27){$\widetilde{A}_n^*$} \put(
10,37){\bf \large = ... = } \hspace{2cm}
\includegraphics[width=45mm]{figure_z-2.eps}
\put(-115,82){$A_2$}\put(-136, 65){$A_3$}
\put(-130,5){$B^*$}\put(-20, 27){$\widetilde{A}_1^*$}
\caption{{\footnotesize The $Z_2$ duality rule.}}
\label{figure
duality rule}
\end{center}
\end{figure}

 Sometimes this duality looks highly non-trivial at the diagram level and requires a
lot of $\gamma$-matrix algebra to convince that the two contribution
are indeed equal. This fact is completely obvious from the
calculations presented in appendix B.

\subsection{The $Z_2$ projection}
\label{Z2 projection subsection} After  evaluation of effective
action (pre-theory) it can be written in the following compact form
\begin{equation}\label{effective action compact form}
S^{eff}_{pre}\ =\ P_a V^a(A)
\end{equation}
Here $P_a$ denote the components of superfield $\mathsf{P}$ and
$V^a(A)$ is a vector field of  the components $A^a$ of the
superfield $\mathsf{A}$, ghosts for SUSY and translations, and
space-time derivatives. Since the action (\ref{effective action
compact form}) is obtained via the integration over a lagrangian
submanifold of BV action, it should satisfy classical BV equation,
which  can be written as\footnote{Solutions of this equation
determine the so-called $\infty$-structure. In the context of
quantum field theories on simplicial complexes these structures were
recently studied in \cite{mnev}. }
\begin{equation}\label{BV equation compact form}
P_bV^a\partial_aV^b\ =\ 0
\end{equation}
In the previous section we discussed the $Z_2$ symmetry of the
effective vertices, which was discovered experimentally at the level
of Feynman diagrams calculation.  Such symmetry in this compact
notations can be written as \begin{equation}\label{symmetry
condition}
\partial_bV^a(A)\eta_{a c}\ =\ \partial_c V^a(A)\eta_{a b}
\end{equation}
where $\eta_{a b}$ is non-degenerate pairing on cohomologies
identifying the components of $\mathsf{A}$ and $\mathsf{P}$ via
$P_a\ =\ \eta_{a b}A^b$ (see \ref{Spectrum of YM(table) 2}). This
condition states that the vertices in the effective theory are
symmetric w.r.t. the interchange of the fields of $\mathsf{A}$ and
$\mathsf{P}$ at the external legs consistent with the $Z_2$ symmetry
of representatives.

Relation (\ref{symmetry condition}) implies that  $V^a$ can be written as a gradient
\begin{equation}\label{solution through the gradient}
V^a(A)\ =\ \eta^{a c}\partial_c \EuScript{F}
\end{equation}
Substituting this solution into BV equation (\ref{BV equation
compact form})
$$
\partial_d\Big(\ \eta^{a c}
\partial_a\EuScript{F}\partial_c\EuScript{F}\ \Big)\ =\ 0
$$
In our calculations the constant of integration can be chosen to be
zero and we come to the conclusion that function $\EuScript{F}$
satisfies classical BV equation on the space of $\mathsf{A}$, namely
$$\eta^{ab}\p_a \EuScript F \p_b \EuScript F  = 0.$$
 The BV
form in this equation coincides with the pairing $\eta_{a b}$
dictated by the $Z_2$ duality on representatives. This function
$\EuScript{F}$ will play the role of BV action, which now depends
only on the component fields of superfield $\mathsf{A}$. Calculation
of the function $\EuScript{F}$ for the pre-action (\ref{pre
effective action}) gives exactly SYM theory coupled to SUSY ghosts.
\be%
\label{effective action}%
\begin{split}
 S^{eff}\ = \int Tr \bigg(-\frac1{1440} F_{\mu\nu}^2 - \frac1{320}
\psi\gamma^\mu D_\mu \psi - A_\rho^* D_\rho c+ gc^*cc +
g[\psi^*,\psi]c  -\\ -\frac32 (\varepsilon\gamma^\mu\psi) A_\mu^* +
\frac1{3} (\varepsilon
\gamma^{\mu\nu}\psi^*)F_{\mu\nu} + \eta^\mu\Big[c^*\p_\mu c -
 A_\nu^*\p_\mu A_\nu  + (\psi^*\p_\mu\psi)\Big] - \ \eta_\mu^*(\e\g^\mu\e)\ - \\
- 40 (\varepsilon\gamma^\mu\varepsilon)(\psi^*\gamma_\mu\psi^*) + 80
(\varepsilon\psi^*)^2 -c^*(\varepsilon\gamma^\mu\varepsilon)A_\mu
\bigg)
\end{split}
\ee%

It is straightforward to check that (\ref{effective action}) satisfy
BV equation. It should be mentioned that at the linear level
(coupling constant $g\ =\ 0$), polynomial $V^a(A)$ have a certain
degree of homogeneity (linear in the fields $A^a$). Hence, the
integration (\ref{solution through the gradient}) needed to extract
action $\EuScript{F}$ from $V^a(A)$ results simply in the factor
$\frac{1}{2}$. Thus at the linear level to obtain the action of the
theory from the action of the pre-theory one should simply identify
the component fields of $\mathsf{A}$ and $\mathsf{P}$ according to
the following rule.

\be%
\label{Spectrum of YM(table) 2}%
\begin{array}{crcl}
\hline
\text{ Polarization} &           A     &       & P \\
\hline 1 & c&&\widetilde{c}\\
 (\lambda\gamma^{\mu}\theta)&A_{\mu}& \ \ \ \ \ \ \ \ \ \ \ \ &\widetilde A_{\mu}(-)\\
 (\lambda\gamma^{\mu}\theta)(\theta\gamma^{\mu})_{\alpha}&\psi^{\alpha}& &\widetilde{\psi}^{\alpha}\\
-16(\lambda\gamma^{\mu}\theta)(\lambda\gamma^{\nu}\theta)(\theta\gamma^{\mu\nu})^{\alpha}&
\psi^*_{\alpha}
&&\widetilde{\psi}^*_\alpha(-)\\
 10(\lambda\gamma^{\mu}\theta)(\lambda\gamma^{\nu}\theta)(\theta\gamma_{\mu\nu\rho}\theta)&A_\rho^*&&\widetilde{A}_\rho^*\\
  (\lambda\gamma^{\mu}\theta)(\lambda\gamma^{\nu}\theta)(\lambda\gamma^{\rho}\theta)
  (\theta\gamma_{\mu\nu\rho}\theta)&c^*&&\widetilde{c}^*(-)\\
\end{array}
\begin{array}{cc}
\ \ \ \ \ \        &  \\
&\widetilde{c} = c^*\\
 &\widetilde A_{\mu} = -A_\mu^*\\
 \Rightarrow&\widetilde{\psi}^{\alpha} = \psi^{\alpha * }\\
&\widetilde{\psi}^*_\alpha = - \psi_\alpha\\
 &\widetilde{A}_\rho^* = A_\rho\\
 &\widetilde{c}^* = -c \\
\end{array}
\ee%
 \begin{picture}(300,-300)(37,0)
\def\axowidth{1.0 }
\SetScale{0.5} \LongArrow(515,43)(620,158)
\LongArrow(515,60)(620,140)
 \LongArrow(515,90)(620,110)

  \LongArrow(515,110)(620,90)
   \LongArrow(515,140)(620,60)
    \LongArrow(515,158)(620,43)
    \LongArrow(620,158)(515,43)
\LongArrow(620,140)(515,60)
 \LongArrow(620,110)(515,90)

  \LongArrow(620,90)(515,110)
   \LongArrow(620,60)(515,140)
    \LongArrow(620,43)(515,158)
\end{picture}%

\vspace {-0.3cm}

 \noindent This rule determines the pairing $\eta_{a
b}$.

\subsubsection{Naive $Z_2$ projection}
Naively, one could expect that just inverting the lines in the
diagrams using the pairing $\eta_{a b}$ would be enough to produce
the BV action of SYM. This naive procedure corresponds to
$$
S^{Naive}\ =\ \eta_{a b} A^a V^b(A)
$$
This however does not solve BV equation. In particular this will
result in the fact that kinetic term for a gauge field would differ
from $ T\!r \, F_{\mu\nu}^2$. This can be checked by explicit
calculation of the diagrams
 (see (\ref{diagr_int1}),(\ref{diagr_int2}) and (\ref{diagr_int3}) in appendix B). The final result for them
 after the identification (\ref{Spectrum of YM(table) 2}) can be written as
$$
L = -\frac1{360} {A}_\mu D_\nu F_{\mu\nu} = -\tfrac1{1440}\Big(
\bold{2}\cdot(\p_{\mu} A_{\nu} - \p_{\nu} A_{\mu})^2 +
\bold3\cdot2g(\p_{\mu} A_{\nu} - \p_{\nu}
A_{\mu})[A_\mu,A_\nu]+\bold4\cdot g^2[A_\mu,A_\nu]^2\Big)
$$
This expression clarifies  that to obtain correct result $ T\!r \,
F_{\mu\nu}^2$ one should put the coefficients $\frac12, \frac13$ and
$\frac14$ in front of the quadratic, cubic and quartic terms
respectively. Remarkably that exactly these coefficients are
dictated  by the procedure (\ref{solution through the gradient}).
Note that this result can not be achieved by rescaling of the
coupling constant $g\rightarrow2/3g$.
%
%

One of the main message of this last section is that the descent of
a symmetry  (for instance SUSY) from the action (\ref{Fund}) down to
(\ref{effective action}) can be realized through the two steps:

\noindent1. One should calculate  the path integral in the
background of the cohomologies  of $Q$ and find the pre-action. It
is important that the action (\ref{Fund}) is the off-shell
supersymmetric version of pre-action.  This descent was discussed in
details in the previous section on the example of the 2-$d$ gauge
model.

\noindent2. One should  implement the procedure (\ref{solution
through the gradient}) to pass from the pre-action to an action.
 For the moment this step can not be done through the path
integral. However, what can be said is that after the application of
the second step the result satisfy classical BV equation over all
the fields including ghosts for SUSY. Hence, some information about
the off-shell  description of SUSY in inherited in the action
(\ref{effective action}).

\section{Acknowledgments}
It is a pleasure to thank N.~Berkovits, S.~Demidov, E.~Ivanov,
C.~Mafra, A.~Morozov, V.~Rubakov, M.~Vasiliev and all the
participants of the Workshop on Pure Spinors (Sao-Paulo, 2006) for
useful discussions. Especially we would like to thank Chris Hull for
pointing out the reference \cite{Hull} to us. We are greatly
indebted to Ulf Gran for providing us the version of his GAMMA
package. This allowed us to speed up considerably the calculations
with gamma matrices for the 10-dimensional Yang-Mills theory. DK and
AL  would like to thank the organizers of the Workshop on Pure
Spinors (Sao-Paulo, 2006)
 for hospitality and creation of the stimulating  atmosphere
 in which a part of this work had been done.
 The work of DK was supported by
 the grant RFBR-04-02-17227,
 the grant of the President of the Russian Federation
NS-7293.2006.2 (government contract 02.445.11.7370) and the
fellowship of Dynasty Foundation in 2007. The work of AL was
supported by the grant RFBR 04-02-17227,  INTAS 03-51-6346 and the
grant for support of scientific schools NSh-8065.2006.2. The work of
VL was supported by the grant RFBR 04-02-16538 and INTAS 03-51-6346.

\newpage

\appendix
\section{Appendix. Some properties of  $SO(10)$ \   $\gamma$ - matrices}
 In this appendix we summarize some properties of $SO(10)$
$\gamma$-matrices, which are important for our calculations. Another
list of useful identities can be found in the appendix of
\cite{Mafra-Berk}. Through the whole paper  we do not distinguish
between upper and lower $SO(10)$ vector indices and use the
convention, that $\gamma^{\mu_1 ... \mu_n} = \frac 1 {n!}
\gamma^{[\mu_1} ... \gamma^{\mu_n]}$.

Ten dimensional  $\gamma$ - matrices can have two upper spinor
indices $(\gamma^{\mu})^{\alpha\beta}$ or two lower indices
$(\gamma^{\mu})_{\alpha\beta}$. Both these two matrices are
symmetric in $\alpha$ and $\beta$.

Using this convention it is straightforward to check  the symmetry
properties of the following representations

\be%
\label{Symmetry propet. for gamma-matr (table)}%
\begin{array}{|c|c|}
\hline
           Symmetric            & Antisymmetric \\
\hline
  (\gamma^{\mu})_{\alpha\beta}\ \ \ \ \ \ \ (\gamma^{\mu_1 ... \mu_4})_{\alpha}{}^{\beta} \
 \ \ \  \ \ (\gamma^{\mu_1 ... \mu_5})_{\alpha\beta}  &     (\gamma^{\mu\nu})_{\alpha}{}^{\beta}   \ \ \
  \ \ \ (\gamma^{\mu\nu\rho})_{\alpha\beta} \\

  (\gamma^{\mu_1 ... \mu_8})_{\alpha}{}^{\beta}
 \ \ \ \ \ \ (\gamma^{\mu_1 ... \mu_9})_{\alpha\beta} &   (\gamma^{\mu_1 ... \mu_6})_{\alpha}{}^{\beta}
 \ \ \ \ \ \
   (\gamma^{\mu_1 ... \mu_7})_{\alpha\beta} \ \ \ \ \ \
   (\gamma^{\mu_1 ... \mu_{10}})_{\alpha}{}^{\beta}\\
 \hline
\end{array}
\ee%
 For the representations with even number of $\gamma$-matrices (which
hence have one upper and one lower spinor index) by symmetry
properties we mean $A_{\alpha}{}^{\beta} = \pm A^{\beta}{}_{\alpha}$

Due to duality properties of $\gamma$-matrices
\be%
\gamma^{\mu_1 ... \mu_n} = \pm \frac1{(10-n)!} \varepsilon^{\mu_1
... \mu_{10}}\ \gamma_{\mu_{n+1} ... \mu_{10}}
\ee%
the basis in the space of all matrices is given by
\be%
\label{basis in gamma-matr}%
\label{complete set for gamma-matr}%
 \delta_{\alpha}{}^{\beta}, \ \ \
(\gamma^{\mu})_{\alpha\beta}, \ \ \  (\gamma^{\mu})^{\alpha\beta}, \
\ \ (\gamma^{\mu\nu})_{\alpha}{}^{\beta}, \ \ \
(\gamma^{\mu\nu\rho})_{\alpha\beta}, \ \ \
(\gamma^{\mu\nu\rho})^{\alpha\beta}, \ \ \ (\gamma^{\mu_1 ...
\mu_4})_{\alpha}{}^{\beta}, \ \ \ (\gamma^{\mu_1 ...
\mu_5})_{\alpha\beta}, \ \ \ (\gamma^{\mu_1 ...
\mu_5})^{\alpha\beta}.
\ee%
The system (\ref{basis in gamma-matr}) is complete. This fact allows
to prove certain Fiertz identities. To illustrate the procedure
consider the following identity
\be%
\label{Fiertz first identity}%
(\gamma^{\mu})_{\alpha\beta}(\gamma_{\mu})_{\delta\sigma} = -\frac1
2 (\gamma^a)_{\alpha\delta}(\gamma_a)_{\beta\sigma} -
\frac1{24}(\gamma^{abc})_{\alpha\delta}
(\gamma_{abc})_{\beta\sigma}.
\ee%
Expanding the l.h.s. in the complete set (\ref{basis in gamma-matr})
with respect to the indices $\alpha$ and $\delta$ one can write
\be%
\label{Fiertz first identity proof}%
(\gamma^{\mu})_{\alpha\beta}(\gamma_{\mu})_{\delta\sigma} = C_1
(\gamma^a)_{\alpha\delta}(\gamma_a)_{\beta\sigma} + C_3
(\gamma^{abc})_{\alpha\delta} (\gamma_{abc})_{\beta\sigma} +
C_5(\gamma^{abcde})_{\alpha\delta} (\gamma_{abcde})_{\beta\sigma}
\ee%
we can use only combinations with odd number of $\gamma$ - matrices
because all spinor indices are lower. Multiplying both sides of
relation (\ref{Fiertz first identity proof}) separately by
$(\gamma^{\nu})^{\beta\delta},(\gamma^{\nu_1\nu_2\nu_3})^{\beta\delta}
$ and  $(\gamma^{\nu_1 ... \nu_5})^{\beta\delta}$ and using the
identities like
\bee%
\label{identities for gamma-matr}%
\begin{split}
&\gamma^a\gamma^{\nu}\gamma_a = -8 \gamma^{\nu}, \\
&\gamma^a\gamma^{\mu\nu\rho}\gamma_a = -4 \gamma^{\mu\nu\rho}, \\
&\gamma^{abc}\gamma^{\mu}\gamma_{abc} = 288 \gamma^{\mu}, \\
&\gamma^{abc}\gamma^{\mu\nu\rho}\gamma_{abc} = -48 \gamma^{\mu\nu\rho}, \\
&\gamma^{abc}\gamma^{\mu_1 ... \mu_5}\gamma_{abc} = 0, \\
&\gamma^{a_1 ... a_5}\gamma^{\mu}\gamma_{a_1 ... a_5} = \gamma^{a_1
... a_5}\gamma^{\mu\nu\rho}\gamma_{a_1 ... a_5} = \gamma^{a_1 ...
a_5}\gamma^{\mu_1 ... \mu_5}\gamma_{a_1 ... a_5} = 0
\end{split}
\eee%
one can straightforwardly  fix the coefficients  $C_1, C_3$ and
$C_5$ in (\ref{Fiertz first identity proof}). In this calculation it
is important to remember the symmetry properties mentioned in the
table (\ref{Symmetry propet. for gamma-matr (table)}). Relations
(\ref{identities for gamma-matr}) can be derived using  the
definition of $\gamma$-matrices: $\{ \gamma^{\mu}, \gamma^{\nu}\} =
2 g^{\mu\nu}$. We acknowledge inestimable help of  Ulf Gran's GAMMA
package \cite{Gran} in doing these computations.

Similar technique allows to obtain other Fiertz identities, for
example
\be%
\label{Fiertz second identity}%
(\gamma^{\mu\nu\rho})_{\alpha\beta}(\gamma_{\mu\nu\rho})_{\delta\sigma}
= -18 (\gamma^a)_{\alpha\delta} (\gamma_a)_{\beta\sigma} + \frac1 2
(\gamma^{abc})_{\alpha\delta} (\gamma_{abc})_{\beta\sigma}
\ee%
{ Independent check that expressions (\ref{Fiertz first identity})
and (\ref{Fiertz second identity}) are consistent can be done by
double expansion of l.h.s of (\ref{Fiertz first identity}): firstly
in the indices $\alpha,\ \delta$, secondly in the indices $\alpha, \
\beta$. Making this double expansion one should come in the end to
the initial expression
$(\gamma^{\mu})_{\alpha\beta}(\gamma_{\mu})_{\delta\sigma}$}.

Applying the same machinery it is straightforward to prove another
useful identity
\be%
\label{identity2}%
(\gamma^{\mu_1 ... \mu_4})_\alpha{}^\beta (\gamma_{\mu_1 ...
\mu_4})_\sigma{}^\delta = 315\  \delta_\alpha{}^\delta
\delta_\sigma{}^\beta + \frac{21}2 (\gamma^{ab})_\alpha{}^\delta
(\gamma_{ab})_\sigma{}^\beta + \frac18 (\gamma^{a_1 ...
a_4})_\alpha{}^\delta(\gamma_{a_1 ... a_4})_\sigma{}^\beta
\ee%
Application of relation (\ref{Fiertz first identity}) allows to
prove useful identity
\be%
\label{Fiertz first identity - consequence}%
(\lambda\gamma^{\mu}\psi)(\lambda\gamma_{\mu}\xi) = 0,\ \ \ \
\forall \psi^{\alpha}, \xi^{\alpha},
\ee%
mentioned in the appendix to \cite{Mafra-Berk}. Indeed, applying
(\ref{Fiertz first identity}) one can find
\be%
\nn%
(\lambda\gamma^{\mu}\psi)(\lambda\gamma_{\mu}\xi) =
 -\frac12 (\lambda\gamma^{\mu}\lambda)(\psi\gamma_{\mu}\xi)  -
\frac1{24}(\lambda\gamma^{abc}\lambda)(\psi\gamma_{abc}\xi) = 0.
\ee%
The first term is equal to zero  due to pure spinor constrains
$(\lambda\gamma^{\mu}\lambda) = 0$, the second one is equal to zero
because $(\gamma^{abc})_{\alpha\beta}$ is antisymmetric in $\alpha,
\beta$ (see table \ref{Symmetry propet. for gamma-matr (table)})
while combination $\lambda^{\alpha}\lambda^{\beta}$ is symmetric.

\hspace{5cm}  \rule{5cm}{0.5pt}

    Sometimes it is useful to have the representation for
    antisymmetric bi-spinor
\be%
\label{bi-spinor antisym}%
\theta^{\alpha}\theta^{\beta} = \frac1{96}
(\theta\gamma^{abc}\theta)(\gamma_{abc})^{\alpha\beta}
\ee%
These relations can be derived in a similar way: for generic
bi-spinor one can write an expansion
\be%
\label{bi-spinor antisym: expansion}%
\xi^{\alpha}\psi^{\beta} = \frac1{16}
(\xi\gamma^{\mu}\psi)(\gamma_{\mu})^{\alpha\beta} +\frac1{96}
(\xi\gamma^{abc}\psi)(\gamma_{abc})^{\alpha\beta} + \frac1{3840}
(\xi\gamma^{\mu_1 ... \mu_5}\psi)(\gamma_{\mu_1 ...
\mu_5})^{\alpha\beta}.
\ee%
The coefficients in this expression can be determined by contracting
both hand sides with $(\gamma^{\mu})_{\alpha\beta},
(\gamma^{\mu\nu\rho})_{\alpha\beta}$ and $(\gamma^{\mu_1 ...
\mu_5})_{\alpha\beta}$. In case of antisymmetric bi-spinor
$\theta\gamma^{\mu}\theta = 0 $ and $\theta\gamma^{\mu_1 ...
\mu_5}\theta = 0 $ because these expressions are symmetric in spinor
indices. The representation for symmetric pure bi-spinor
gives\footnote{We are indebted to Carlos Mafra for the correction of
the coefficient in this expression.}
\be%
\label{bi-spinor symmetr}%
\lambda^{\alpha}\lambda^{\beta} = \frac1{3840}(\lambda \gamma^{\mu_1
... \mu_5}\lambda)(\gamma_{\mu_1 ... \mu_5})^{\alpha\beta}
\ee%
due to constrains $(\lambda\gamma^{\mu}\lambda) = 0$.

In our calculations we will also need a representation
\cite{Mafra-Berk}

\be%
\label{identity from appendix}%
(\lambda\gamma^{\mu} \theta) (\lambda\gamma^{\nu} \theta) =
\frac12(\lambda\gamma^{a \mu\nu} \theta) (\lambda\gamma_a \theta),
\ee%
which can be proven by applying formula (\ref{Fiertz first
identity}) and commutation relations $\{ \gamma^{\mu},
\gamma^{\nu}\} = 2 g^{\mu\nu}$ in the r.h.s.

One more useful formula \cite{Berk-Nekr} is
\be%
\label{identity from appendix 2}%
(\gamma^{\mu\nu})_{\alpha}{}^{\delta}(\gamma_{\mu\nu})_{\beta}{}^{\sigma}
= -8 \delta_{\alpha}{}^{\sigma} \delta_{\beta}{}^{\delta} + 4
(\gamma^{\mu})_{\alpha\beta} (\gamma_{\mu})^{\delta\sigma} -
2\delta_{\alpha}{}^{\delta}\delta_{\beta}{}^{\sigma}
\ee%
In our calculations we also use scalar\footnote{$
\delta^{\mu\nu\rho}_{abc} = \frac1{3!}\delta^{[\mu}_a \delta^\nu_b
\delta^{\rho]}_c$} product \cite{Mafra-Berk}
\bee%
\label{scalar product1}%
 \ll\!\!(\lambda\gamma^{\mu}\theta)(\lambda\gamma^{\nu}\theta)(\lambda\gamma^{\rho}\theta)(\theta\gamma_{abc}\theta)\!\!\gg
 = \frac1{120} \delta^{\mu\nu\rho}_{abc}\\
 \label{scalar product2}%
  \ll\!\!(\lambda\gamma^{\mu\nu\rho}\theta)(\lambda\gamma_{p}\theta)(\lambda\gamma_{q} \theta)(\theta\gamma_{abc}\theta)\!\!\gg
 = \frac1{70} \delta^{[\mu}_{[p}\eta_{q][a}^{}\delta^\nu_b \delta^{\rho ]}_{c]}
\eee%

%
%
%
%
%
%
%
%

\section{Appendix. \ Calculation of effective action for the $D=10$ SYM}
\label{section - app2} In this appendix we present the calculation
of effective action \cite{we} for the theory
\be%
\label{fund SYM}%
S^{SUSY} = \int\ Tr\Big( <\EuScript{P},\  Q_{B} \EuScript{A}> \ +\
g <\EuScript{P},\  \EuScript{A}^2\!>\ + \ <\EuScript{P},\ \e^\a
Q^s_\a \EuScript{A}>\ + \ \!<\EuScript{P},\ \eta^\mu P^{s}_\mu
\EuScript{A}>\  -\ \ \eta_\mu^*(\e\g^\mu\e) \Big)
\ee%
 along the way described in section 5, namely  using  the scalar product $\ll\ \gg$,
defined in (\ref{scalar product1}) and (\ref{scalar product2}).
Operators $Q_B,\ P_s$ and $Q_s$ are defined as
\begin{equation}
 Q_B \ =Q + \Phi = \  \lambda^\alpha \frac{\partial}{\partial
\theta^\alpha}+ \frac12 \theta^\alpha \frac{\partial f^\mu}{\partial
\lambda^\alpha} \partial_\mu,\ \ \ \  \e^\a Q^s_\a \ =\  \e^\a
\frac{\p}{\p \t^\a} - (\e\g^\mu\t) \frac{\p}{\p x^\mu} , \ \ \
P^s_\mu \ =\ \frac{\p}{\p x^\mu}
\end{equation}

Polarizations of  component fields  are given
 by (\ref{Spectrum of YM(appendix)}) (see (\ref{Spectrum of YM(table)}))

\be%
\label{Spectrum of YM(appendix)}%
\begin{array}{cccc}
\hline
\text{ Polarization} &           A     &       & P \\
\hline 1 & c& &\widetilde{c}\\
 (\lambda\gamma^{\mu}\theta)&A_{\mu}&&\widetilde A_{\mu}\\
 (\lambda\gamma^{\mu}\theta)(\theta\gamma^{\mu})_{\alpha}&\psi^{\alpha}&&\widetilde{\psi}^{\alpha}\\
-16(\lambda\gamma^{\mu}\theta)(\lambda\gamma^{\nu}\theta)(\theta\gamma^{\mu\nu})^{\alpha}&
\psi^*_{\alpha}
&&\widetilde{\psi}^*_\alpha\\
 10(\lambda\gamma^{\mu}\theta)(\lambda\gamma^{\nu}\theta)(\theta\gamma_{\mu\nu\rho}\theta)&A_\rho^*&&\widetilde{A}_\rho^*\\
  (\lambda\gamma^{\mu}\theta)(\lambda\gamma^{\nu}\theta)(\lambda\gamma^{\rho}\theta)
  (\theta\gamma_{\mu\nu\rho}\theta)&c^*&&\widetilde{c}^*\\

\end{array}
\ee%

 \noindent We
start from the quadratic terms in the action and then derive
interaction terms.

\subsubsection *{ Quadratic level} First of all we evaluate the
diagrams which do not depend on the ghosts $\e^\a$ and $\eta^\mu$.
These diagrams contribute into the classical part of effective
action. Below we list the nontrivial diagrams, intermediate and
final expressions for them. The first diagram is

\begin{center}
{\includegraphics[width=30mm]{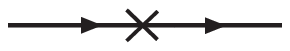}}
\put(-70,-5){$c$}\put(-20,-7){$\widetilde{A}_\mu$}\put(-48,15){$\Phi$}
\end{center}

We emphasize that  one should  first act by all the operators and
propagators onto the incoming  field  and then
 project the result onto cohomologies according the procedure
 discussed in section 5. Namely, to project the result $R$ onto a representative $h$
  corresponding to a certain component field $B$ one should
  calculate $\bl\ \ \ \br$  with the complementary representative $d$, definition is given in (\ref{Z2 duality
  on representatives}). The contribution into the action will be given by
 $ (-1)^{ \#\! \widetilde{B}} \bl d\widetilde{B} \cdot R\br $. Here $\#\! \widetilde{B}$ denotes the parity of the component field $ \widetilde{B}$.
 The relative sign $ (-1)^{ \#\! \widetilde{B}} $ will be taken into account only in the final result (\ref{PRE effective lagrangian SYM}).

 The contribution of the first diagram is given by
\be%
\label{diagr1}%
 \EuScript{D} = \bl 10 (\l\g_\mu\t)(\l\g_\nu\t)(\t\g^{\mu
\nu \rho}\t) \widetilde{A}_\rho \cdot (\l\g_a\t)\p_a c \br = -
\widetilde{A}_\rho {10 \over 120} \d^{\mu \nu a}_{\mu \nu \rho} \p_a
c = - \widetilde{A}_\rho\p_\rho c.
\ee%
Here we used the parities of component fields and the definition
(\ref{scalar product1}).

 \vspace{-5pt} \hspace{5cm}
 \rule{5cm}{0.5pt}

Dual diagram is
\begin{center}
{\includegraphics[width=30mm]{figure_app1.eps}}
\put(-70,-5){$A^*_\mu$}\put(-20,-5){$\widetilde{c}^*$}\put(-48,15){$\Phi$}
\end{center}
\be%
\label{diagr2}%
\EuScript{D} =  \bl  \widetilde{c}^* \cdot (\l\g^a\t) \p_a \cdot 10(\l\g_\mu\t)
(\l\g_\nu\t) (\t\g^{\mu \nu \rho}\t) A_\rho^* \br = - {10 \over
120} \widetilde{c}^*\d^{\mu \nu a} _{\mu \nu \rho}  \p_a A_\rho^* =
- A_\rho^*\p_\rho \widetilde{c}^*.
\ee%

\vspace{-5pt} \hspace{5cm}  \rule{5cm}{0.5pt}

\begin{center}
{\includegraphics[width=30mm]{figure_app1.eps}}
\put(-70,-5){$\psi$}\put(-20,-7){$\widetilde{\psi}^*$}\put(-48,15){$\Phi$}
\end{center}
\be%
\begin{split}
\label{diagr3}%
\EuScript{D} = \bl  (\l\g_a\t)(\t\g_a\widetilde{\psi}^*)
(\l\g^\nu\t)(\l\g^\mu\t)(\t\g^{\mu}\p_\nu \psi) \br = -\tfrac{1}{96}
 \bl (\l\g_a\t)(\l\g_\nu\t) (\l\g_\mu\t)(\t\g^{b c
d}\t) \br
 \widetilde{\psi}^* \g^a \g_{b c d }\g^\mu\p_\nu \psi = \\ = - {1
\over 96\cdot120}\widetilde{\psi}^*\g^a \g_{a\nu \mu }\g^\mu \p_\nu
\psi =- {72 \over 96\cdot 120} \widetilde{\psi}^* \g_\nu\p_\nu \psi
= -{1 \over 160} \widetilde{\psi}^* \g_\nu\p_\nu \psi.
\end{split}
\ee%
Here we used (\ref{bi-spinor antisym}) and the definition of $\g$ -
matrices. This diagrams give kinetic term for the fermion.

\vspace{-5pt} \hspace{5cm}  \rule{5cm}{0.5pt}

\begin{center}
{\includegraphics[width=50mm]{figure_app2.eps}} \put(-130,-5){$A_\mu
$}\put(-20,-5){$\widetilde{A}^*_\mu$}\put(-103, 14){$\Phi$}\put(-50,
14){$\Phi$}
\end{center}
This diagram is responsible for the kinetic term for the gauge
field. First of all we calculate the action  of operator~$\Phi$
$$
\Phi (\l \g^\mu \t) A_\mu = (\l \g^\mu \t) (\l \g^\nu \t) \p_\mu
A_\nu = \frac12 (\l \g^{a \mu \nu}\t) (\l \g_a \t)\p_\mu A_\nu.
$$
In the last transformation we used (\ref{identity from appendix}).
This should be done to prepare the result for application of the
propagator (it is
 impossible to do it directly on $(\l \g^\mu \t) (\l \g^\nu \t)$
because $(\g^\mu)_{\a \b}$ is symmetric w.r.t. $\a$ and $\b$). The
pre-mage is given by
\be%
\label{kinetic term fo gauge} \Pi= \frac14 (\t \g^{a \mu \nu} \t
)(\l \g_a \t) \p_\mu A_\nu.\ee%
 Indeed, $Q\cdot\Pi = \l {\p \over \p\t} \Pi=
\frac12 (\l \g^{a \mu \nu} \t )(\l \g_a \t) \p_\mu A_\nu$ due to
pure spinor constraints $(\l \g_a\l) = 0$. The whole contribution of
the diagram is given by
\be%
\label{diagr4}%
\EuScript{D} = \bl \frac14
(\l\g_\rho\t)\widetilde{A}^*_\rho (\l\g_\e\t) (\t\g^{a \mu \nu }\t) (\l\g^a\t)\br  \p_\e
\p_\mu A_\nu  = \tfrac1{4\cdot120}\d^{\rho \e a} _{a \mu \nu }
\widetilde{A}^*_\rho \p_\e \p_\mu A_\nu = -
{1\over360}(\widetilde{A}^*_\mu \p^2 A_\mu - \widetilde{A}^*_\mu \p_\mu \p_\nu A_\nu ).\ee%
 If one identifies  $\widetilde{A}^*_\mu = A_\mu$ it is possible  recognize  in this result abelian part of ${1 \over 720} F^2_{\mu
\nu}$.

 \vspace{-5pt} \hspace{5cm}
\rule{5cm}{0.5pt}

Now we switch to the calculation  of diagrams proportional to the
ghost for the SUSY. The first diagram is

\begin{center}
{\includegraphics[width=30mm]{figure_app1.eps}} \put(-70,-5){$A_\mu
$}\put(-20,-7){$\widetilde{\psi}$} \put(-50,15){$\e Q_s$}
\end{center}
It is enough  to use only the second part of $Q_s$, proportional to
space-time derivative\footnote{The other term of $Q_s$ does not
contribute to the final result of the diagram due to the degree in
$\l$ and $\t$. \label{footnote app}}
$$\e Q_s (\l \g^\mu \t) A_\mu = - (\e \g^\nu
\t) (\l \g^\mu \t)\p_\nu A_\mu.$$ The contribution of the diagram is
\be%
\label{diagr5}%
\EuScript{D} = 16 \bl(\l \g^a \t)(\l \g^b \t)(\t \g^{a b}
\widetilde{\psi})(\e \g^\nu \t)( \l \g^\mu \t)\br \p_\nu A_\mu =
\!\tfrac{1}{6} \bl(\l \g^a \t)(\l \g^b \t)(\l \g^\mu \t)(\t \g^{mnk}
\t)\br\times \\ \times(\e\g^\nu\g^{mnk}\g^{ ba}\widetilde{\psi}
)\p_\nu A_\mu \nn   =- {1 \over 6\cdot120}(\e\g^\nu\g^{a b \mu}\g^{a
b}\widetilde{\psi} )\p_\nu A_\mu = {1\over10}(\e \g^\nu
\g^\mu\widetilde{\psi})\p_\nu A_\mu
.\ee%

 \vspace{-5pt}
\hspace{5cm} \rule{5cm}{0.5pt}

 The $Z_2$ dual diagram is
\vspace{-5pt}
\begin{center}
{\includegraphics[width=30mm]{figure_app1.eps}} \put(-70,-5){$\psi^*
$}\put(-20,-7){$\widetilde{A}^*_\mu$} \put(-50,15){$\e Q_s$}
\end{center}
Application of supercharge to $\psi^*$ gives
$$ \e Q_s [-16(\l \g^\mu
\t)(\l \g^\nu \t)(\t \g^{\mu\nu} \psi^*)] = 16(\e \g^\rho \t)(\l
\g^\mu \t)(\l\g^\nu \t)(\t\g^{\mu \nu}\p_\rho\psi^*)$$ The
contribution of the diagram is
\be%
\label{diagr6}%
\EuScript{D} = 16\bl(\l \g^\rho\t)\widetilde{A}^*_\rho
(\e\g^\l\t)(\l\g^\mu\t)(\l\g^\nu\t)(\t\g^{\mu\nu}\p_\l\psi^*)\br
 =
-{1\over10}(\e\g^\mu\g^\nu\p_\mu\psi^*)\widetilde{A}^*_\nu
.\ee%

\vspace{-5pt} \hspace{5cm}  \rule{5cm}{0.5pt}

\begin{figure}[h]
\begin{center}
{\includegraphics[width=30mm]{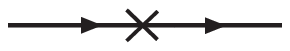}} \put(-70,-5){$\psi
$}\put(-20,-7){$\widetilde{A}_\mu$} \put(-50,15){$\e Q_s$}
\end{center}\caption{}
\label{figure of diagr7}
\end{figure}
In contrast to previous pair of diagrams in this case it is enough
to use only the first term${}^{\ref{footnote app}}$ in $Q_s$ acting
as derivative $\p \over\p\t$
$$
\e Q_s[(\l\g^\mu\t)(\t\g^\mu\psi)] =
(\l\g^\mu\e)(\t\g^\mu\psi)-(\l\g^\mu\t)(\e\g^\mu\psi).
$$
Using the formula (\ref{Fiertz first identity}) one can expand the
contribution of the first term and obtain

\be%
\begin{split}\label{diagr7}%
\EuScript{D} = 10
\bl(\l\g^a\t)(\l\g^b\t)(\t\g^{ab\rho}\t)\widetilde{A}_\rho(\l\g^\mu\e)(\t\g^\mu\psi)\br\
- \ 10 \bl
(\l\g^a\t)(\l\g^b\t)(\t\g^{ab\rho}\t)\widetilde{A}_\rho(\l\g^\mu\t)\br(\e\g^\mu\psi)
= \\ = -\frac{10}2  \ll\!\! (\l\g^a\t)(\l\g^b\t)(\t\g^{a b
\rho}\t)(\l\g^\mu\t)\!\!\gg (\e\g^{\mu}\psi)\widetilde{A}_\rho -
\frac{10}{24} \ll\!\! (\l\g^a\t)(\l\g^b\t)(\t\g^{a b
\rho}\t)(\l\g^{\mu\nu c}\t)\!\!\gg (\e\g^{\mu \nu
c}\psi)\widetilde{A}_\rho \\ - 10  \ll\!\!
(\l\g^a\t)(\l\g^b\t)(\t\g^{a b \rho}\t)(\l\g^\mu\t)\!\!\gg
(\e\g^{\mu}\psi)\widetilde{A}_\rho  =
-{10\over2\cdot120}\d^{ab\mu}_{ab\rho}(\e\g^\mu\psi)\widetilde{A}_\rho
- \\ - {10\over120}
\d^{ab\mu}_{ab\rho}(\e\g^\mu\psi)\widetilde{A}_\rho
 - {10\over24\cdot70}\delta^{[\mu}_{[a}\eta_{b][a}^{}\delta^\nu_b
\delta_{\rho ]}^{c]}(\e\g_{\mu \nu c}\psi)\widetilde{A}_\rho =
-\frac3{2} (\e\g_{\rho}\psi)\widetilde{A}_\rho .\end{split}
\ee%
The last term with the coefficient $\frac{10}{24\cdot70}$ is equal
to zero due to symmetry properties.

\vspace{-5pt} \hspace{5cm}  \rule{5cm}{0.5pt}

Calculation of dual diagram is more tricky.
\begin{figure}[H]
\begin{center}
{\includegraphics[width=30mm]{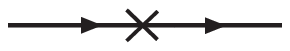}}
\put(-70,-5){$A^*_\mu $}\put(-20,-7){$\widetilde{\psi}^*$}
\put(-50,15){$\e Q_s$}
\end{center}\caption{}
\label{figure of diagr8}
\end{figure}
\noindent Again operator $Q_s$ acts as derivative${}^{\ref{footnote
app}}$ $\p\over\p\t$
\be%
\label{diagr8.1}%
\e Q_s[ 10 (\l\g^\mu\t) (\l\g^\nu\t) (\t\g^{\mu\nu\rho} \t
)A_\rho^*] = 20(\l\g^\mu\e) (\l\g^\nu\t)(\t\g^{\mu \nu\rho}\t)
A_\rho^* + 20(\l\g^\mu\t) (\l\g^\nu\t)(\e\g^{\mu \nu\rho}\t)
A_\rho^*
. \ee%
Evaluating the contribution of the first term one can find
\be%
\bl
(\l\g^a\t)(\t\g^a\widetilde{\psi}^*)(\l\g^\mu\e)(\l\g^\nu\t)(\t\g^\mu\g^\nu\g^\rho\t)\br
= - \frac12
\bl(\l\g^a\t)(\t\g^a\widetilde{\psi}^*)(\l\g^\nu\t)(\l\g^b\t)(\e\g^b\g^\nu\g^\rho\t)
\br  -\nn \\ -\nn
 \frac1{24}
\bl(\l\g^a\t)(\t\g^a\widetilde{\psi}^*)(\l\g^\nu\t)(\l\g^{mnp}\t)(\e\g_{mnp}\g^\nu\g^\rho\t)
\br\  =   - {1 \over 2\cdot96\cdot120}(\e\g^b\g^\nu\g^\rho\g_{a \nu
b}\g^a\widetilde{\psi}^*) - \\ - \tfrac1{24\cdot96\cdot70}
\delta^{[m}_{[a}\eta_{\e][p}^{}\delta^n_q \delta^{k ]}_{c]} (\e
\g_{mnk}\g^\e\g^\rho\g^{pqc}\g^a\widetilde{\psi}^*) = \Big( -
\tfrac{432}{2\cdot96\cdot120} - \tfrac{1008}{24\cdot96\cdot70}
\Big)(\e\g^\rho\widetilde{\psi}^*) =
 - \frac1{40}(\e\g^\rho\widetilde{\psi}^*).
\nn \ee%
Here we used that $\t\g^{\mu\nu\rho}\t = \t\g^\mu\g^\nu\g^\rho\t $
(This is due to  $\t\g^\mu\t = 0 $) , identity (\ref{Fiertz first
identity}) and scalar products (\ref{scalar product1}),(\ref{scalar
product2}). The second term in (\ref{diagr8.1}) gives
$$
\bl
(\l\g^a\t)(\t\g^a\widetilde{\psi}^*)(\l\g^\mu\t)(\l\g^\nu\t)(\e\g^{\mu\nu\rho}\t)\br
= \frac1{96\cdot120}(\widetilde{\psi}^*\g^a\g_{a
\mu\nu}\g^{\mu\nu\rho}\e) = -
\frac1{20}(\e\g^\rho\widetilde{\psi}^*)
$$
Collecting together
\be%
\label{diagr8.2}%
\EuScript{D} = -20 \cdot \frac1{40}
(\e\g^\rho\widetilde{\psi}^*)A_\rho^* - 20\cdot \frac1{20}
(\e\g^\rho\widetilde{\psi}^*)A_\rho^* =
-\frac3{2}(\e\g^\rho\widetilde{\psi}^*)A_\rho^* .
\ee%
Even from this calculation it is clear that the $Z_2$ duality
discussed in section 5 looks highly non-trivial at the level of
Feynman diagrams. Calculation of the diagram in fig. \ref{figure of
diagr7} is considerably simpler than that in the fig. \ref{figure of
diagr8}. However, the final result after the identification (\ref{Spectrum of YM(table) 2}) is the same. In the next set of
diagrams we will see more dramatic realization of this duality.
Calculation from the one side of this duality looks very simple,
calculation from the other side  requires a lot of $\g$-matrix
algebra.

\vspace{-5pt} \hspace{5cm}  \rule{5cm}{0.5pt}

\begin{center}
{\includegraphics[width=50mm]{figure_app2.eps}} \put(-130,-5){$A_\mu
$}\put(-20,-5){$\widetilde{\psi}$}\put(-105, 14){$\e Q_s$}\put(-50,
14){$\Phi$}
\end{center}
Application of sypersymmetry operator${}^{\ref{footnote app}}$ gives
$Q$-exact expression
$$
\e Q_s[(\l\g^\mu\t)A_\mu] = (\l\g^\mu\e)A_\mu.
$$
The pre-image is $(\t\g^\mu\e)A_\mu$. The whole contribution gives
\be%
\label{diagr9}%
\EuScript{D} = -16 \bl
(\l\g^a\t)(\l\g^b\t)(\t\g^{ab}\widetilde{\psi})(\l\g^\nu\t)(\t\g^\mu\e)\br
\p_\nu A_\mu  =-
\tfrac{16}{96\cdot120}(\widetilde{\psi}\g^{ab}\g_{ab\nu}\g^\mu\e)\p_\nu
A_\mu =  \frac1{10}(\e\g^\mu\g^\nu\widetilde{\psi})\p_\nu A_\mu
\ee%

\vspace{-5pt} \hspace{5cm}  \rule{5cm}{0.5pt}

\begin{center}
{\includegraphics[width=50mm]{figure_app2.eps}}
\put(-130,-5){$\psi^*$}\put(-20,-7){$\widetilde{A}^*_\mu$}\put(-105,
14){$\e Q_s$}\put(-50, 14){$\Phi$}
\end{center}
$$
\e Q_s [-16 (\l\g^\mu\t)(\l\g^\nu\t)(\t\g^{\mu\nu}\psi^*)] = -2\cdot
16(\l\g^\mu\e)(\l\g^\nu\t)(\t\g^{\mu\nu}\psi^*) - 16
(\l\g^\mu\t)(\l\g^\nu\t)(\e\g^{\mu\nu}\psi^*)
$$
The second term is proportional to that one which appeared in the
calculation of  kinetic term for the gauge field, and according to
(\ref{kinetic term fo gauge})
\be%
\label{app1}%
\Pi_2 = -4 (\t \g^{a \mu \nu} \t )(\l \g_a \t)(\e \g^{\mu\nu}\psi^*)
 \ee%
 To write the first term in the convenient
form one should expand $\g^{\mu\nu} = g^{\mu\nu} - \g^{\nu}\g^{\mu}$
and use (\ref{Fiertz first identity - consequence})
\be%
\label{app2}
 \Pi_1 = 32 (\t\g^\mu\e)(\l\g^\nu\t)(\t\g^\nu\g^\mu\psi^*)
\ee%
$$
\Phi(\Pi_1+\Pi_2) =32 (\l\g^\rho\t)(\t\g^\mu\e)(\l\g^\nu\t)
(\t\g^\nu\g^\mu\p_\rho \psi^*) - 4(\l\g^\rho\t)(\t\g^{a\mu\nu}\t)
(\l\g_a\t)(\e\g^{\mu\nu}\p_\rho \psi^*)
$$

\begin{multline}
\label{diagr10}%
\EuScript{D} =  32 \bl
(\l\g^d\t)\widetilde{A}^*_d(\l\g^\rho\t)(\t\g^\mu\e)(\l\g^\nu\t)
(\t\g^\nu\g^\mu\p_\rho \psi^*)\br  - 4 \bl (\l\g^d\t)
\widetilde{A}^*_d (\l\g^\rho\t)(\t\g^{a\mu\nu}\t)
(\l\g_a\t)\br(\e\g^{\mu\nu}\p_\rho \psi^*) = \\ =
-\frac{32}{96\cdot120}(\e\g^\mu
\g_{d\rho\nu}\g^\nu\g^\mu\p_\rho\psi^*)\widetilde{A}^*_d
-\frac4{120} \widetilde{A}^*_d \d^{a\mu\nu}_{d\rho
a}(\e\g^{\mu\nu}\p_\rho\psi^*) = - \frac2{9}
(\e\g^{\nu\mu}\psi^*)\p_\nu \widetilde{A}^*_\mu
\end{multline}

\vspace{-5pt} \hspace{5cm}  \rule{5cm}{0.5pt}

\begin{center}
{\includegraphics[width=50mm]{figure_app2.eps}}
\put(-130,-5){$A_\mu$}\put(-20,-5){$\widetilde{\psi}$}\put(-105,
14){$\Phi$}\put(-50, 14){$\e Q_s$}
\end{center}

The first part of the diagram is analogous to the one responsible
for the kinetic term in the gauge field (\ref{kinetic term fo
gauge}) $ \Pi= \frac14 (\t \g^{a \mu \nu} \t )(\l \g_a \t) \p_\mu
A_\nu $. Applying  operator $\e Q_s$ one can  come to following
contribution
$$
\EuScript{D} = -\frac{16}2 \bl
(\l\g^b\t)(\l\g^c\t)(\t\g^{bc}\widetilde{\psi})(\e
\g^{a\nu\mu}\t)(\l\g^a\t)\br \p_\nu A_\mu - \frac{16}4\bl
(\l\g^b\t)(\l\g^c\t)(\t\g^{bc}\widetilde{\psi})(\t
\g^{a\nu\mu}\t)(\l\g^a\e)\br \p_\nu A_\mu
$$
Contribution of the first and second terms respectively
$$
\EuScript{D}_1 = \tfrac{16}{2\cdot96\cdot120} (\e\g^{a
\nu\mu}\g_{bca}\g^{bc}\widetilde{\psi})\p_\nu A_\mu =
\tfrac{576\cdot 16}{2\cdot96\cdot120 }(\e
\g^{\mu\nu}\widetilde{\psi})\p_\nu A_\mu = \frac2{5}(\e
\g^{\mu\nu}\widetilde{\psi})\p_\nu A_\mu
$$
$$
\EuScript{D}_2 =
\tfrac{16}{4\cdot2\cdot96\cdot120}(\widetilde{\psi}\g^{bc}\g_{bca}\g^{\mu\nu}\g^a\e)\p_\nu
 A_\mu - \tfrac{16}{4\cdot24\cdot96\cdot70}\delta^{[a}_{[b}\eta_{c][k}^{}\delta^q_l \delta^{r
 ]}_{p]}(\widetilde{\psi}\g^{bc}\g_{klp}\g^{\mu\nu}\g_{aqr}\e)\p_\nu A_\mu =
  \frac1{15}(\e \g^{\mu\nu}\widetilde{\psi})\p_\nu A_\mu
$$
Finally
\be%
\label{diagr10}%
\EuScript{D} = \frac{7}{15}(\e \g^{\mu\nu}\widetilde{\psi})\p_\nu
A_\mu
\ee%

 \vspace{-5pt} \hspace{5cm}  \rule{5cm}{0.5pt}

\begin{center}
{\includegraphics[width=50mm]{figure_app2.eps}}
\put(-130,-5){$\psi^*$}\put(-20,-5){$\widetilde{A}^*_\mu$}\put(-105,
14){$\Phi$}\put(-50, 14){$\e Q_s$}
\end{center}
Applying operator $\Phi$ and taking the  pre-image one obtains
$$
\Pi = 4 (\t
\g^{a\rho\nu}\t)(\l\g_a\t)(\l\g^\mu\t)(\t\g^\mu\g^\nu\p_\rho\psi^*)
$$
Application of $\e Q_s$ gives 4 terms, corresponding contributions
are
$$
\EuScript{D}_1 = 8 \widetilde{A}^*_b \bl (\l\g^b\t) (\e
\g^{a\rho\nu}\t)(\l\g_a\t)(\l\g^\mu\t)(\t\g^\mu\g^\nu\p_\rho\psi^*)
\br  = \frac2{5} (\e\p_\rho\psi^*)\widetilde{A}^*_\rho
-\frac4{15}(\e \g^{b\rho}\p_\rho\psi^*)\widetilde{A}^*_b
$$
$$
\EuScript{D}_2 = 4 \widetilde{A}^*_b\bl (\l\g^b\t) (\t
\g^{a\rho\nu}\t)(\l\g_a\e)(\l\g^\mu\t)(\t\g^\mu\g^\nu\p_\rho\psi^*)
\br   = \frac1{10} (\e\p_\rho\psi^*)\widetilde{A}^*_\rho -
\frac1{30}(\e \g^{b\rho}\p_\rho\psi^*)\widetilde{A}^*_b
$$
$$
\EuScript{D}_3 = -4\widetilde{A}^*_b \bl (\l\g^b\t) (\t
\g^{a\rho\nu}\t)(\l\g_a\t)(\l\g^\mu\e)(\t\g^\mu\g^\nu\p_\rho\psi^*)
\br  = -\frac1{5} (\e\p_\rho\psi^*)\widetilde{A}^*_\rho
$$
$$
\EuScript{D}_4 = 4 \widetilde{A}^*_b \bl (\l\g^b\t) (\t
\g^{a\rho\nu}\t)(\l\g_a\t)(\l\g^\mu\t)(\e\g^\mu\g^\nu\p_\rho\psi^*)
\br  = -\frac2{5} (\e\p_\rho\psi^*)\widetilde{A}^*_\rho
-\frac2{45}(\e \g^{b\rho}\p_\rho\psi^*)\widetilde{A}^*_b
$$
The total contribution is
\be%
\label{diagr11}%
\EuScript{D} = -\frac{1}{10}(\e\p_\rho\psi^*)\widetilde{A}^*_\rho -
\frac{31}{90}(\e \g^{\mu\nu}\psi^*)\p_\mu \widetilde{A}^*_\nu
\ee%

\noindent Again  we emphasize that the $Z_2$ duality  after the identification (\ref{Spectrum of YM(table) 2}) implies the diagram
identity
\begin{center}
$$
\begin{array}{ccc}
{\includegraphics[width=50mm]{figure_app2.eps}}
\put(-130,-5){$A_\mu$}\put(-20,-5){$\widetilde{\psi}$}\put(-105,
14){$\Phi$}\put(-50, 14){$\e Q_s$} &&
{\includegraphics[width=50mm]{figure_app2.eps}}
\put(-130,-5){$\psi^*$}\put(-20,-5){$\widetilde{A}^*_\mu$}\put(-105,
14){$\e Q_s$}\put(-50, 14){$\Phi$}\\ +& \ \ \ = \ \ \  &+
\\{\includegraphics[width=50mm]{figure_app2.eps}}
\put(-130,-5){$A_\mu$}\put(-20,-5){$\widetilde{\psi}$}\put(-105,
14){$\e Q_s$}\put(-50,
14){$\Phi$}&&{\includegraphics[width=50mm]{figure_app2.eps}}
\put(-130,-5){$\psi^*$}\put(-20,-5){$\widetilde{A}^*_\mu$}\put(-105,
14){$\Phi$}\put(-50, 14){$\e Q_s$}
\end{array}
$$
\end{center}
Which is rather non-trivial!

\vspace{-5pt} \hspace{5cm}  \rule{5cm}{0.5pt}

\begin{center}
\includegraphics[width=50mm]{figure_app2.eps}
\put(-130,-5){$\psi^*$}\put(-20,-5){$\widetilde{\psi}$}\put(-105,
14){$\e Q_s$}\put(-50, 14){$\e Q_s$}
\end{center}
 The calculation of the first part of this diagram is analogous to
 (\ref{app1}),(\ref{app2}). Application of the second operator $\e Q_s (\Pi_1
 +\Pi_2)$ gives 5 terms.
 $$
 \EuScript{D}_1 = -2\cdot 16^2 \bl
 (\l\g^d\t)(\l\g^k\t)(\t\g^{dk}\widetilde{\psi})(\e\g^\mu\e)(\l\g^\nu\t)(\t\g^\nu\g^\mu\psi^*)\br
 = 32(\e\g^\mu\e)(\widetilde{\psi}\g^\mu\psi^*)
 $$
  $$
 \EuScript{D}_2 = 2\cdot 16^2 \bl
 (\l\g^d\t)(\l\g^k\t)(\t\g^{dk}\widetilde{\psi}(\t\g^\mu\e)(\l\g^\nu\e)(\t\g^\nu\g^\mu\psi^*)\br
 = -\tfrac85(\widetilde{\psi}\g^a\g^\mu\e)(\e\g^a\g^\mu\psi^*)+
 $$
 $$+\tfrac1{45}(\widetilde{\psi}\g^{abc\mu}\e)(\e\g^{abc\mu}\psi^*)
 +\tfrac8{15}(\widetilde{\psi}g^{bc}\e)(\e\g^{bc}\psi^*)
 $$
  $$
 \EuScript{D}_3 = -2\cdot 16^2 \bl
 (\l\g^d\t)(\l\g^k\t)(\t\g^{dk}\widetilde{\psi})(\t\g^\mu\e)(\l\g^\nu\t)(\e\g^\nu\g^\mu\psi^*)\br
 = -\tfrac{16}{5}(\widetilde{\psi}\g^a\g^\mu\e)(\e\g^a\g^\mu\psi^*)
 $$
  $$
 \EuScript{D}_4 = \frac12 \cdot 16^2\bl (\l\g^d\t)(\l\g^k\t)(\t\g^{dk}\widetilde{\psi})(\e\g^{a
 \mu\nu}\t)(\l\g_a\t)(\e\g^{\mu\nu}\psi^*)\br = - \tfrac{32}{5}(\widetilde{\psi}\g^{bc}\e)(\e\g^{bc}\psi^*)
 $$
  $$
 \EuScript{D}_5 = \frac14 \cdot 16^2\bl (\l\g^d\t)(\l\g^k\t)(\t\g^{dk}\widetilde{\psi})(\t\g^{a
 \mu\nu}\t)(\l\g_a\e)(\e\g^{\mu\nu}\psi^*)\br  = -\tfrac{16}{15}(\widetilde{\psi}\g^{bc}\e)(\e\g^{bc}\psi^*)
 $$
Summing up all the contributions and using the identities
(\ref{identity2}) and (\ref{identity from appendix 2}) one can come
to
\be%
\label{diagr12}%
\EuScript{D} =  80(\e\g^\mu\e)(\widetilde{\psi}\g^\mu\psi^*)
-160(\e\widetilde{\psi})(\e\psi^*)
\ee%

\vspace{-5pt} \hspace{5cm}  \rule{5cm}{0.5pt}

\begin{center}
\includegraphics[width=50mm]{figure_app2.eps}
\put(-130,-5){$A_\mu$}\put(-20,-5){$\widetilde{c}$}\put(-105,
14){$\e Q_s$}\put(-50, 14){$\e Q_s$}
\end{center}
Applying $\e Q_s$ one obtains an exact expression.
$$
\e Q_s[(\l\g^\mu\t)A_\mu] = (\l\g^\mu\e)A_\mu, \ \ \Longrightarrow\
\ \Pi = (\t\g^\mu\e)A_\mu.
$$
The whole contribution is
\be%
\label{diagr13}%
\EuScript{D} =  \widetilde{c}\bl
(\l\g^\mu\t)(\l\g^\nu\t)(\l\g^\rho\t)(\t\g_{\mu\nu\rho}\t)(\e\g^a\e)A_a\br
=\widetilde{c}(\e\g^\mu\e)A_\mu
\ee%

\vspace{-5pt} \hspace{5cm}  \rule{5cm}{0.5pt}

\begin{center}
\includegraphics[width=50mm]{figure_app2.eps}
\put(-130,-5){$c^*$}\put(-20,-5){$\widetilde{A}^*$}\put(-105,
14){$\e Q_s$}\put(-50, 14){$\e Q_s$}
\end{center}
The calculation of dual diagram is much more tricky. Operator $Q_s$
acts as $\e \frac{\partial}{\partial\theta}$ and gives 2 terms \bee
 \e\frac{\partial}{\partial\theta}
[(\lambda\gamma^\mu\theta)(\lambda\gamma^\nu\theta)(\lambda\gamma^\rho\theta)(\theta\gamma^{\mu\nu\rho}\theta)c^\ast]=
\nn\\
\nn3(\lambda\gamma^\mu\e)(\lambda\gamma^\nu\theta)(\lambda\gamma^\rho\theta)(\theta\gamma^{\mu\nu\rho}\theta)c^\ast
-2(\lambda\gamma^\mu\theta)(\lambda\gamma^\nu\theta)(\lambda\gamma^\rho\theta)(\e\gamma^{\mu\nu\rho}\theta)c^\ast,
\eee next step

\be%
\Pi_1=3(\theta\gamma^\mu\varepsilon)(\lambda\gamma^\nu\theta)(\lambda\gamma^\rho\theta)(\theta\gamma^{\mu\nu\rho}\theta)c^\ast
\nn \\ \nn \Pi_2 =
-\frac{1}{2}(\theta\gamma^{\mu\nu\sigma}\theta)(\lambda\gamma_\sigma\theta)(\lambda\gamma^\rho\theta)(\varepsilon\gamma^{\mu}\gamma^{\nu}\gamma^\rho\theta)c^\ast,
\ee%
After applying second $\e Q_s$ there arise seven terms
\be%
\e Q_s(\Pi_1 +\Pi_2) =
3(\varepsilon\gamma^\mu\varepsilon)(\lambda\gamma^\nu\theta)(\lambda\gamma^\rho\theta)(\theta\gamma^{\mu\nu\rho}\theta)
-6(\theta\gamma^\mu\varepsilon)(\lambda\gamma^\nu\varepsilon)(\lambda\gamma^\rho\theta)(\theta\gamma^{\mu\nu\rho}\theta) \nn \\
-6(\theta\gamma^\mu\varepsilon)(\lambda\gamma^\nu\theta)(\lambda\gamma^\rho\theta)(\varepsilon\gamma^{\mu\nu\rho}\theta)
-(\varepsilon\gamma^{\mu\nu\sigma}\theta)(\lambda\gamma_\sigma\theta)(\lambda\gamma^\rho\theta)(\varepsilon\gamma^{\mu}\gamma^{\nu}\gamma^\rho\theta)\nn \\
-\frac{1}{2}(\theta\gamma^{\mu\nu\sigma}\theta)(\lambda\gamma_\sigma\varepsilon)(\lambda\gamma^\rho\theta)(\varepsilon\gamma^{\mu}\gamma^{\nu}\gamma^\rho\theta)
+\frac{1}{2}(\theta\gamma^{\mu\nu\sigma}\theta)(\lambda\gamma_\sigma\theta)(\lambda\gamma^\rho\varepsilon)(\varepsilon\gamma^{\mu}\gamma^{\nu}\gamma^\rho\theta)\nn\\
-\frac{1}{2}(\theta\gamma^{\mu\nu\sigma}\theta)(\lambda\gamma_\sigma\theta)(\lambda\gamma^\rho\theta)(\varepsilon\gamma^{\mu}\gamma^{\nu}\gamma^\rho\varepsilon)
\nn \ee%

Some calculus  gives us the following results \bee \bl
3(\varepsilon\gamma^\mu\varepsilon)(\lambda\gamma^\nu\theta)(\lambda\gamma^\rho\theta)(\theta\gamma^{\mu\nu\rho}\theta)
(\lambda\gamma^a\theta)\br&=&\tfrac{3}{10}(\varepsilon\gamma^a\varepsilon)
,\nn\\
 -6\bl
(\theta\gamma^\mu\varepsilon)(\lambda\gamma^\nu\theta)(\lambda\gamma^\rho\theta)(\varepsilon\gamma^{\mu\nu\rho}\theta)
(\lambda\gamma^a\theta)\br&=&\tfrac{9}{40}(\varepsilon\gamma_{
a}\varepsilon)
,\nn\\
 -\bl
(\varepsilon\gamma^{\mu\nu\sigma}\theta)(\lambda\gamma_\sigma\theta)(\lambda\gamma^\rho\theta)(\varepsilon\gamma^{\mu}\gamma^{\nu}\gamma^\rho\theta)
(\lambda\gamma^a\theta)\br&=&\tfrac{1}{4}(\varepsilon\gamma_{
a}\varepsilon)
,\nn\\
-\tfrac{1}{2}\bl(\theta\gamma^{\mu\nu\sigma}\theta)(\lambda\gamma_\sigma\theta)(\lambda\gamma^\rho\theta)(\varepsilon\gamma^{\mu}\gamma^{\nu}\gamma^\rho\varepsilon)(\lambda\gamma^a\theta)\br&=&\tfrac{1}{10}(\varepsilon\gamma^a\varepsilon)
,\nn\\
-6\bl(\theta\gamma^\mu\varepsilon)(\lambda\gamma^\nu\varepsilon)(\lambda\gamma^\rho\theta)(\theta\gamma^{\mu\nu\rho}\theta)(\lambda\gamma^a\theta)\br&=&(\tfrac{3}{32}-\tfrac{3}{160})(\varepsilon\gamma^a\varepsilon)
,\nn\\
-\tfrac{1}{2}\bl(\theta\gamma^{\mu\nu\sigma}\theta)(\lambda\gamma_\sigma\varepsilon)(\lambda\gamma^\rho\theta)(\varepsilon\gamma^{\mu}\gamma^{\nu}\gamma^\rho\theta)(\lambda\gamma^a\theta)\br&=&
(\tfrac{13}{320}-\tfrac{1}{64})(\varepsilon\gamma^a\varepsilon)
,\nn\\
\tfrac{1}{2}
\bl(\theta\gamma^{\mu\nu\sigma}\theta)(\lambda\gamma_\sigma\theta)(\lambda\gamma^\rho\varepsilon)(\varepsilon\gamma^{\mu}\gamma^{\nu}\gamma^\rho\theta)(\lambda\gamma^a\theta)\br&=&(
\tfrac{1}{20}-\tfrac{1}{40} )(\varepsilon\gamma^a\varepsilon). \nn
\eee Finally we obtain the same result as in (\ref{diagr13})
\be%
\label{diagr14}%
\EuScript{D} =  c^*(\e\g^\mu\e)\widetilde{A}^*_\mu
\ee%

\vspace{-5pt} \hspace{5cm}  \rule{5cm}{0.5pt}

\begin{center}

$$
\begin{array}{ccccc}
{\includegraphics[width=30mm]{figure_app1.eps}}
\put(-70,-5){$c$}\put(-20,-5){$\widetilde{c}$}\put(-52,15){$\eta^\mu\p_\mu$}
&\ \ + \ \ \  & {\includegraphics[width=30mm]{figure_app1.eps}}
\put(-70,-5){$c^*$}\put(-20,-5){$\widetilde{c}^*$}\put(-52,15){$\eta^\mu\p_\mu$}
\\ && \\
\label{diagr16 trans}%
\EuScript{D} =  \widetilde{c}\eta^\mu \p_\mu c
&&
\label{diagr16 trans}%
\EuScript{D} =  c^*\eta^\mu \p_\mu
\widetilde{c}^*

\end{array}
$$
\end{center}

\hspace{5cm} \rule{5cm}{0.5pt} \vspace{-5pt}

\begin{center}

$$
\begin{array}{ccccc}
{\includegraphics[width=30mm]{figure_app1.eps}}
\put(-70,-5){$A_\mu$}\put(-20,-7){$\widetilde{A}_\mu$}\put(-52,15){$\eta^\mu\p_\mu$}
&\ \ + \ \ \  & {\includegraphics[width=30mm]{figure_app1.eps}}
\put(-70,-5){$A_\mu^*$}\put(-20,-7){$\widetilde{A}^*_\mu$}\put(-52,15){$\eta^\mu\p_\mu$}
\\&&\\
\EuScript{D} = \widetilde{A}_\mu\eta^\nu \p_\nu A_\mu  &&
  \EuScript{D} =
A_\mu^*\eta^\nu \p_\nu \widetilde{A}^*_\mu
\end{array}
$$
\end{center}

\vspace{-5pt} \hspace{5cm}  \rule{5cm}{0.5pt}

\begin{center}

$$
\begin{array}{ccc}
{\includegraphics[width=30mm]{figure_app1.eps}}
\put(-70,-5){$\psi$}\put(-20,-6){$\widetilde{\psi}$}\put(-52,15){$\eta^\mu\p_\mu$}
&\ \ + \ \ \  & {\includegraphics[width=30mm]{figure_app1.eps}}
\put(-70,-5){$\psi^*$}\put(-20,-6){$\widetilde{\psi}^*$}\put(-52,15){$\eta^\mu\p_\mu$}
\\&&\\
\EuScript{D} = \eta^\mu (\widetilde{\psi}\p_\mu\psi) &&
\EuScript{D} = \eta^\mu (\psi^*\p_\mu\widetilde{\psi}^*)
\end{array}
$$
\end{center}

\subsubsection*{Interaction terms}
The next step is to calculate  the contributions  proportional to
the gauge coupling constant $g$. In contrast to the model discussed
in section \ref{2d gauge model}, the only effect of this interaction
is the replacement of operator  insertion $\Phi =
(\l\g^\mu\t)\p_\mu$ by the gauge field $(\l\g^\mu\t)A_\mu$. In this
calculation it is important to remember that  all possible
permutations of external  legs must be  considered to obtain the
correct gauge invariant result for the effective action. For
example, consider the diagram
\begin{center}
\includegraphics[width=50mm]{figure_app2.eps}
\put(-130,-5){$A_\mu$}\put(-20,-5){$\widetilde{A}_\mu^*$}\put(-105,
14){$\Phi$}\put(-50, 14){$\Phi$}
\end{center}

one should add five diagrams

\begin{center}
{\includegraphics[width=50mm]{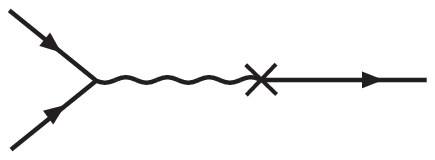}}
\put(-125,42){$A_\mu$}\put(-125,-0){$A_\mu$} \put(-20,
13){$\widetilde{A}^*_\mu$}\put(-58, 34){$\Phi$}
\end{center}

\be%
\label{diagr_int1}%
\EuScript{D} = g\tfrac1{360}\widetilde{A}^*_\mu\p_\nu[A_\mu,A_\nu]
\ee%

\begin{center}
$$
\begin{array}{ccc}
{\includegraphics[width=50mm]{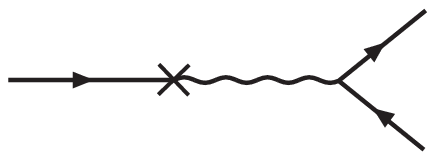}}
\put(-120,13){$A_\mu$}\put(-27,0){$A_\mu$} \put(-27,
42){$\widetilde{A}^*_\mu$}\put(-92, 34){$\Phi$} &\ \ \ \ \  &
{\includegraphics[width=50mm]{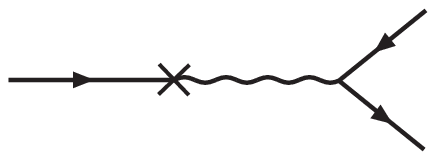}}
\put(-120,13){$A_\mu$}\put(-27,0){$\widetilde{A}^*_\mu $} \put(-27,
42){$A_\mu$}\put(-92, 34){$\Phi$}
\end{array}
$$
\end{center}
\be%
\label{diagr_int2}%
\EuScript{D} = g\tfrac1{360}[\widetilde{A}^*_\mu A_\nu] (\p_\mu
A_\nu -\p_\nu A_\mu)
\ee%

\begin{center}
$$
\begin{array}{ccc}
{\includegraphics[width=40mm]{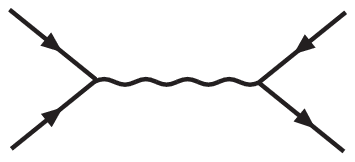}}
\put(-100,-0){$A_\mu$}\put(-27,0){$\widetilde{A}^*_\mu$} \put(-27,
45){${A}_\mu$}\put(-100, 43){$A_\mu$} &\ \ \ \ \  &
{\includegraphics[width=40mm]{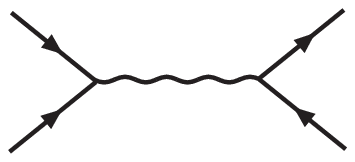}}
\put(-100,-0){$A_\mu$}\put(-27,0){$A_\mu$} \put(-27,
45){$\widetilde{A}^*_\mu$}\put(-100, 43){$A_\mu$}
\end{array}
$$
\end{center}

\be%
\label{diagr_int3}%
\EuScript{D} = g^2\tfrac1{360}[\widetilde{A}^*_\mu A_\nu]
[A_\mu,A_\nu]
\ee%

which are different due to the clockwise rule (see section 5 of
\cite{we}).

\vspace{-5pt} \hspace{5cm}  \rule{5cm}{0.5pt}

There is also a subset of trivial diagrams  with ghosts. These
diagrams are depicted below.

\begin{center}

$$
\begin{array}{ccccc}
{\includegraphics[width=33mm]{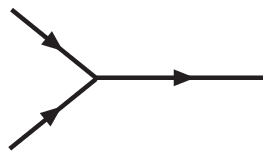}}
\put(-80,50){$\phi^*$}\put(-80,0){$\phi$} \put(-27,
15){$\widetilde{c}^*$} &\ \ \ \ \  &
{\includegraphics[width=33mm]{figure_app5.eps}}
\put(-80,50){$c$}\put(-80,0){$\phi$} \put(-27,
15){$\widetilde{\phi}$} &\ \ \ \ \  &
{\includegraphics[width=33mm]{figure_app5.eps}}
\put(-80,50){$c$}\put(-80,0){$\phi^*$} \put(-27,
15){$\widetilde{\phi}^*$}
\end{array}
$$
\end{center}
Here $\phi$ stand for arbitrary field ($c, A_\mu, \psi$). It is
important to remember that all the permutations of external legs
should be taken into account according to the clockwise rule.

Collecting together the results for all the diagrams one can come to
the following result for the effective lagrangian.

\be%
\begin{split}
 \label{PRE effective lagrangian SYM}%
L_{pre}^{eff} = - \frac1{360} \widetilde{A}^*_\mu D_\nu F_{\mu\nu} +
\frac1{160} \psi\gamma^\mu D_\mu \widetilde{\psi}^*  + \frac g{160}
\widetilde{A}_\mu^*(\psi\gamma^\mu \psi)   + \widetilde{A}_\mu D_\mu
c +  A^*_\mu D_\mu \widetilde{c}^* -  g A^*_\mu [
\widetilde{A}_\mu^*,c]
  +
  \\ + g \big( \widetilde{c}cc + \widetilde{c}^*[c^*,c]  +
  [\widetilde{\psi}^*,\psi^*]c +
[\widetilde{\psi},\psi]c + \widetilde{c}^*[\psi^*,\psi]\big) +
\\
 +\frac3{2} (\varepsilon\gamma^\mu\psi) \widetilde{A}_\mu
 +\frac3{2} (\varepsilon\gamma^\mu\widetilde{\psi}^*) A_\mu^*
  + \frac2{3}
(\varepsilon \gamma^{\mu\nu}\widetilde{\psi})D_{\mu}A_\nu +\frac2{3}
(\varepsilon \gamma^{\mu\nu}\psi^*)D_{\mu}\widetilde{A}^*_\nu -\frac
g{3} (\varepsilon \gamma^{\mu\nu}\widetilde{\psi})[A_{\mu},A_\nu]+
\\ + \eta^\mu\Big[\widetilde{c}\p_\mu c - c^*\p_\mu \widetilde{c}^* +  \widetilde{A}_\nu\p_\mu A_\nu -
 A_\nu^*\p_\mu \widetilde{A}^*_\nu  +
(\widetilde{\psi}\p_\mu\psi) - (\psi^*\p_\mu\widetilde{\psi}^*)\Big]
-\eta^*_\mu (\e\g^\mu\e)-
\\
- 80
(\varepsilon\gamma^\mu\varepsilon)(\widetilde{\psi}\gamma_\mu\psi^*)
+ 160 (\varepsilon\widetilde{\psi})(\varepsilon\psi^*)
-c^*(\varepsilon\gamma^\mu\varepsilon)\widetilde{A}^*_\mu
-\widetilde{c}(\varepsilon\gamma^\mu\varepsilon)A_\mu
\end{split}
\ee%
Here $D_\mu = \p_\mu + g[A_\mu, \cdot]$, $F_{\mu\nu} = \p_\mu A_\nu
- \p_\nu  A_\mu + g [A_\mu,A_\nu]$.

\end{document}